\documentclass{aa}  

\usepackage{natbib}
\bibpunct{(}{)}{;}{a}{}{,}
\usepackage{graphicx}
\usepackage{float}
\usepackage{subcaption}
\usepackage[normalem]{ulem}
\usepackage{txfonts}
\usepackage{amsmath}
\usepackage{svg}
\usepackage{longtable}
\usepackage{multicol}
\usepackage{afterpage}
\usepackage{placeins}
\usepackage{bm}
\usepackage{dblfloatfix}
\usepackage[all]{nowidow}
\usepackage[colorlinks=true, linkcolor=black, urlcolor=black, citecolor=blue]{hyperref}

\begin{document} 

   \title{Enhancing the detection of low-energy M dwarf flares: Wavelet-based denoising of CHEOPS data}

   \subtitle{}

   \author{J. Poyatos
          \inst{1,2} \href{https://orcid.org/0000-0003-4064-4268}{\protect\includegraphics[height=0.25cm]{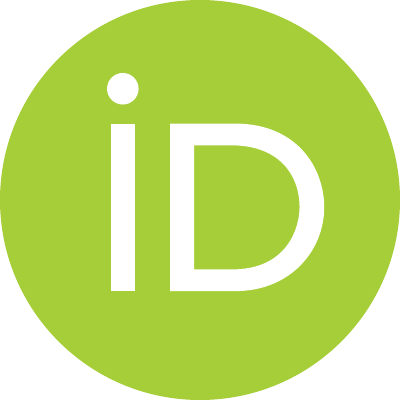}}
          \and
          O. Fors\inst{1,2} \href{https://orcid.org/0000-0002-4227-9308}{\protect\includegraphics[height=0.25cm]{Fig/orcid.pdf}}
          \and
          J.M. Gómez Cama\inst{3,2} \href{https://orcid.org/0000-0003-0173-5888}{\protect\includegraphics[height=0.25cm]{Fig/orcid.pdf}}
          \and
          I. Ribas\inst{2,4} \href{https://orcid.org/0000-0002-6689-0312}{\protect\includegraphics[height=0.25cm]{Fig/orcid.pdf}}
          }

   \institute{Departament de Física Quàntica i Astrofísica, Institut de Ciències del Cosmos (ICCUB), Universitat de Barcelona (IEEC-UB),
Martí i Franquès 1, E-08028 Barcelona, Spain\\
\email{julienpoyatos@icc.ub.edu}
        \and
            Institut d'Estudis Espacials de Catalunya (IEEC), 08860 Castelldefels (Barcelona), Spain
         \and
             Departament d’Enginyeria Electrònica i Biom{\`e}dica, Institut de Ciències del Cosmos (ICCUB), Universitat de Barcelona, IEEC-UB, Martí i Franquès 1,
E-08028 Barcelona, Spain
        \and
            Institut de Ci{\`e}ncies de l’Espai (ICE, CSIC), Campus UAB, c/de Can Magrans s/n, E-08193 Bellaterra, Barcelona, Spain
             }

   \date{Received 19 December 2024; accepted 29 May 2025}

 \abstract
   {Stellar flares are powerful bursts of electromagnetic radiation that are triggered by magnetic reconnection in the chromosphere of stars. They occur frequently and intensely on active M dwarfs. While missions such as TESS and Kepler have studied regular and superflares, their detection of flares with energies below $10^{30}$ erg remains incomplete. An extension of flare studies to include these low-energy events could enhance flare formation models and provide insight into their impact on exoplanetary atmospheres.}
   {This study investigates the capacity of CHEOPS to detect low-energy flares in M dwarf light curves. Using the high photometric precision and observing cadence of CHEOPS, along with a tailored wavelet-based denoising algorithm, we improved the detection completeness and refined flare statistics for low-energy events.}
   {We conducted a flare injection and recovery process to optimise the denoising parameters, applied it to the CHEOPS light curves to maximise flare detection rates, and used a flare-breakdown algorithm to analyse complex structures.}
   {Our analysis recovered 291 flares with energies ranging from $3.7 \times 10^{26}$ to $8.9 \times 10^{30}$ erg for 62 M dwarfs, about $\sim$42\% of which exhibited complex, multi-peaked structures. The denoising algorithm improved the flare recovery by $\sim$35\%, although it marginally extended the lower boundary of detectable energies. For the full sample, the power-law index $\alpha$ was $1.99 \pm 0.10$, but a log-normal distribution fitted better. This suggests multiple possible flare-formation scenarios.}
   {While the observing mode of CHEOPS is not ideal for large-scale surveys, it captures weaker flares than TESS and Kepler, and thus extends the observed energy range. Wavelet-based denoising enhances the recovery of low-energy events, which enables us to explore the micro-flaring regime. The expansion of low-energy flare observations could refine flare-generation models and improve our understanding of their role in star-planet interactions.}

   \maketitle

\section{Introduction}

Stellar flares are stochastic and intense bursts of electromagnetic radiation and charged particles that are triggered by magnetic reconnection in the chromosphere of stars. This phenomenon occurs when the magnetic field lines in the chromosphere realign, which releases energy across the electromagnetic spectrum. This can generate flares of varying intensity and duration and can significantly increase the stellar luminosity \citep{Benz_2010}. 
The stellar magnetic activity significantly influences the frequency and energy of stellar flares. M dwarfs are known for their high levels of magnetic activity and for flaring more frequently than earlier-type stars because their dynamo processes are more vigorous and yield a higher incidence of magnetic reconnection events \citep{Kowalski_2024}. These stars can experience more frequent and energetic flares that sometimes reach energy levels that are several orders of magnitude more intense than solar flares \citep{Bruevich_2001, Howard_2018}.
The effects of stellar flares on planetary habitability are various. On the one hand, stellar flares can damage planetary atmospheres: X-ray and ultraviolet radiation, along with charged particles that are emitted during flares, can deplete atmospheric layers, especially on planets with weak magnetic fields or located near their host star. This atmospheric erosion can compromise the potential for habitability by reducing the protection against harmful radiation and altering the planetary climate \citep{Ribas_2016, Airapetian_217}. On the other hand, the energy that is radiated during stellar flares may play a role in triggering prebiotic chemical processes: The increased ultraviolet radiation during flares can trigger chemical reactions that are crucial for the emergence of life. This suggests that flares may be necessary for prebiotic chemistry to happen on planets orbiting M dwarfs \citep{Scalo_2007, Rimmer_2018}.

It is essential to understand the formation mechanisms, frequency, and energy distribution of stellar flares for several reasons. It provides insights into stellar magnetic activity and its variation with spectral type, especially in comparison to our Sun. By investigating the flare activity of various stellar types, it is additionally possible to refine models of stellar and planetary evolution \citep{Gudel_2007, Lammer_2012}. TESS (Transiting Exoplanet Survey Satellite) and Kepler have provided valuable data on flare occurrences with cadences of minutes to a dozen seconds. The 20-second cadence of TESS revealed that complex flare structures are frequent, including multi-peak profiles and quasi-periodic pulsations (QPPs). These complex structures suggest that these non-classical flare shapes may result from emission cascades within a single active region or from sympathetic flares from neighbouring regions \citep{Davenport_2016, Howard_2022}. In this context, the study of smaller-scale flare events becomes increasingly important for a better characterisation of stellar variability and the formation mechanisms of flares. Extending photometric precision and observing cadence down to a few seconds could greatly improve our understanding of these phenomena and might reveal even finer flare structures. This increased sensitivity may enable us to detect micro- and nano-flares, which are small-scale flaring events with energies ranging from $10^{22}$ to $10^{27}$ erg that are thought to play a crucial role in coronal heating \citep{Parker_1988}. No concluding evidence supports micro- and nano-flares against other possible explanations, however, mainly because it is hard to gather precise statistics about these flares because they are hidden in the instrumental noise. It is therefore still unclear whether micro- and nano-flares follow the same trends as more powerful flaring events. Furthermore, flare formation scenarios that diverge from self-organised criticality propose that flares might consist of an avalanche of several combined micro- and nano-flares. This would have implications for the current understanding of the impact of flares on exoplanetary atmospheres \citep{Audard2000, Sheikh_2016, Aschwanden_2019}.

The space telescope CHaracterising ExOPlanet Satellite (CHEOPS) is primarily designed to measure exoplanet transits. It offers exceptional photometric precision. With a precision of approximately 20 ppm over a 6-hour integration period for a V$\sim$9 star and down to 150 ppm for shorter timescales, CHEOPS surpasses other photometric instruments in the visible spectrum, such as TESS and current ground-based telescopes \citep{Benz_2021, Oddo_2023}. The exposure time of CHEOPS can also reach 0.001 seconds for very bright stars (V$\sim$6), which is several orders of magnitude faster than those of TESS and Kepler \citep{Borucki_2010, Ricker_2015}. Although CHEOPS is limited to on-target observations, it provides high photometric precision and a fast accessible observing cadence. These are crucial for capturing the short and stochastic nature of low-energy flares.

To enhance the detection of flares in the CHEOPS light curves, denoising techniques based on the wavelet transform can filter out noise from the signal. The wavelet transform decomposes time-series data into different frequency components, which facilitates the isolation and reduction of noise. Unlike the Fourier transform, which only provides frequency characteristics of the signal, the wavelet transform recovers both temporal and frequency information, making it more efficient in analysing transient signals such as stellar flares. The discrete wavelet transform (DWT) is commonly used as a base for time-series denoising algorithms, and it has many applications outside of astronomy \citep{Pasti_1999, Polat_2018, Jang_2021}. The main downside of the DWT is that it introduces a downsampling of the input signal by definition, which can lead to a loss of details due to its non-redundant nature \citep{Dohono_1994}. In contrast, the stationary wavelet transform (SWT) avoids this by upsampling the wavelet coefficients at each decomposition level. Although this involves a higher computational cost, the SWT provides the advantage of maintaining the original length of the signal at each decomposition level and providing a redundant, shift-invariant representation of the data \citep{Rhif_2019, Kumar_2021}. In the context of astronomy, techniques based on the SWT have been developed for noise filtering and demonstrated significant improvements in the signal-to-noise ratios (S/N) and the ability to recover faint astrophysical signals \citep{Strack_2006}. These methods emphasise the importance of adapting the current wavelet-based denoising techniques to the specific characteristics of astronomical time-series and show that these denoising algorithms could enhance the detection of faint transient phenomena such as low-energy flares. 

An efficient wavelet-based denoising algorithm is crucial for optimising the extraction of meaningful flare statistics in our analysis pipeline using CHEOPS data. We explore the performance of the combination of the photometric precision of CHEOPS and an SWT-based denoising algorithm in enhancing the flare detection in M dwarf light curves. In Section \ref{section:methods} we describe the target selection, the data reduction, the SWT parameter exploration, the flare detection, and flare-breakdown algorithms. In Section \ref{section:results} we quantify the improvement in flare recovery that is achieved by denoising, we analyse the obtained flare catalogues, and we compare our results to the literature. In Section \ref{section:discussion} we discuss the limitations of this study and suggest several avenues for improvement. We conclude in Section \ref{section:conclusion}.
 
\section{Methods}
\label{section:methods}

\subsection{Target selection and data reduction}
\label{section:target_selection}

Among the data gathered by CHEOPS since the start of its operations, several observational programmes have targeted late-type stars. CH\_PR100018 (PI I. Pagano) is a prominent programme that has been actively investigating the variability of late-K and M dwarf stars since the beginning of scientific observations in April 2020. This programme is particularly focused on characterising the photometric variability of these stars, including flares and star spots.
CH\_PR100010 (PI. G Szabó) operated from July 2020 to December 2023. While its primary aim was to study the debris disks around various spectral types, this programme included observations of several late-type stars to explore the interaction between stellar activity and circumstellar dust. 
CH\_PR130057 (PI G. Szabó) was another significant programme, running from August 2022 to September 2023, and specifically targeted the young M dwarf star AU Mic. Known for its strong stellar activity and prominent debris disk, AU Mic is an ideal subject for examining the correlation between stellar flares and its surrounding environment. 

\begin{figure}
\centering
\includegraphics{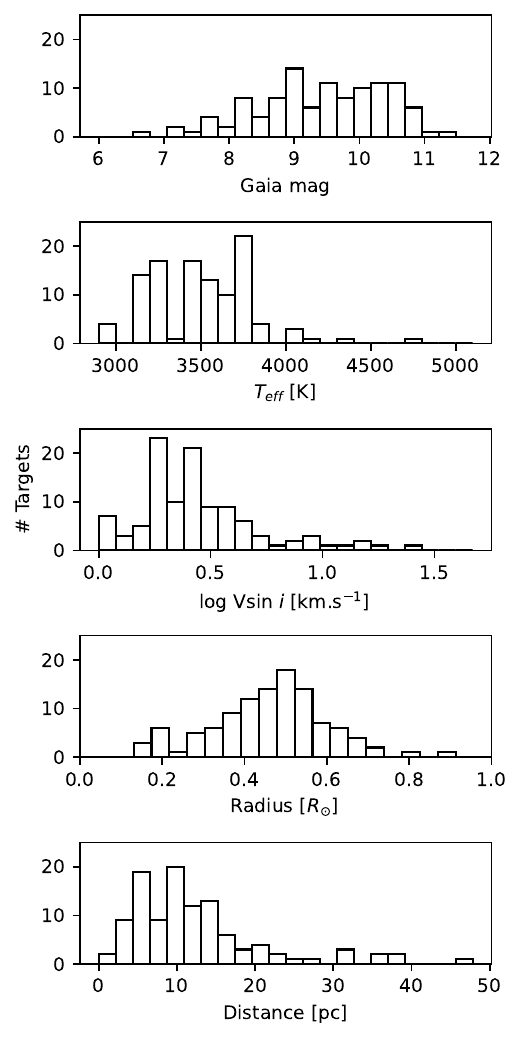}
\caption{Parameters of our stellar sample. From top to bottom: Histograms for the Gaia magnitude, $T_{\mathrm{eff}}$, log($v$\,sin\,$i$), radius, and distance.}
\label{fig:star_description}
\end{figure}

The nominal observation mode of CHEOPS produces subarrays (200-pixel diameter) every few tens of seconds by co-adding multiple short sub-exposures onboard. In parallel, smaller cutouts called imagettes (60-pixel diameter) are downlinked at a higher cadence and can be used to construct light curves with finer temporal resolution. While individual sub-exposures can be as short as 0.001 seconds, the fastest cadence for downlinked subarrays is 22.65 seconds, set by telemetry and onboard processing constraints. In contrast, imagettes can be downlinked with cadences as short as 2.8 seconds. All programmes mentioned above adopted a 3-second cadence for imagettes to balance S/N, avoid saturation, and maximise duty cycle \citep{Boldog_2023, Bruno_2024}. This cadence remains well-suited for detecting and analysing rapid, transient events like stellar flares and offers higher time resolution than the 20-second cadence of TESS.
To take advantage of the high precision and fast observing cadence provided by CHEOPS, we retrieved all publicly available imagettes data on main sequence M dwarfs from the CHEOPS archive\footnote{\url{https://cheops-archive.astro.unige.ch/archive_browser/}}. Due to the proprietary period associated with CHEOPS programmes, imagettes data are not yet available for all observations. We therefore completed our sample with the raw light curves provided in \cite{Bruno_2024}. In total, we obtained 66.65 days of on-target time from the three programmes, comprising observations of 110 different M dwarfs. For each target, we retrieved the Gaia \textit{G}-band magnitude ($G_{\mathrm{mag}}$) and effective temperature ($T_{\mathrm{eff}}$) from the CHEOPS file metadata. We searched for $H_{\alpha}$ equivalent width values from \cite{Schofer_2019}, as well as log $R'_{HK}$ values from \cite{Astudillo_2017} and \cite{Boro_2018}, but found $H_{\alpha}$ EW values for only 64 of the 110 stars and log $R'_{HK}$ values for 71 stars. This limited availability prevented us from providing a comprehensive activity indicator distribution for the entire sample. Instead, we used the mean rotational velocity ($V$\,sin\,$i$) values from Table A.1 of \cite{Bruno_2024} as an alternative indicator of stellar activity. We calculated the distance to each target using its parallax from Gaia DR3, or Gaia DR2 if not available. We obtained the radius of each star from the TESS Input Catalogue. The parameter distributions in our sample are shown in Figure \ref{fig:star_description}, and the full list is available in Table~\ref{target_list}. As expected from the focus of the three observing programmes, our sample is heavily biased towards AU Mic, with more than 340 visits observed, while all other stars have been observed during less than 70 visits each. 

The CHEOPS Data Reduction Pipeline (\texttt{DRP}) does not automatically process imagettes datasets \citep{Hoyer_2020}. We hence relied on \texttt{PIPE}\footnote{\url{https://github.com/alphapsa/PIPE}}, a photometric extraction package developed by the CHEOPS consortium to complement the official \texttt{DRP}, to reduce imagettes and extract Point Spread Function (PSF) photometry \citep{Brandeker_2024}. We optimised the PSF parameters for each target by allowing the software to freely explore the impact of the radius of the fitting region, the number of components for the fit, a simultaneous fit for the background, dark current correction, and static image removal through a gradient-like search algorithm. In the case of the light curves recovered by \cite{Bruno_2024}, this imagettes reduction was already performed using an ad hoc tool, the Data Reduction Tool (\texttt{DRT}), described in \cite{Morgado_2022} and \cite{Fortier_2024}. We consistently found marginal differences of less than 50 ppm when comparing raw light curves extracted through both \texttt{PIPE} and the \texttt{DRT}.

\subsection{Detrending}
Variability in fast-cadence light curves can arise from multiple sources \citep{Sulis_2020, Kalman_2025}. In the case of CHEOPS, white noise primarily results from photon counting, while time-correlated variations originate from both instrumental and astrophysical effects. To mitigate instrumental systematics, we used the package \texttt{PyCHEOPS}\footnote{\url{https://github.com/pmaxted/pycheops}} \citep{Maxted_2022, Maxted_2023}, a tool developed by the CHEOPS consortium to remove correlations between flux and various instrumental parameters. Specifically, we removed linear, quadratic, and cubic flux correlations with time, centroid position, sine and cosine of the roll angle, background flux, contamination, and smear. 
Among astrophysical contributors, granulation in M dwarfs is expected to be negligible at CHEOPS precision \citep{Sulis_2023}. The dominant source of residual variability can thus be attributed to surface inhomogeneities associated with magnetic active regions (such as spots, faculae, pores, etc) that induce rotational modulation. To account for this, we applied a Savitzky-Golay (Savgol) filter with a 20-minute window, corresponding to the duration of the longest flare detected in \cite{Bruno_2024}. This approach effectively removes trends on timescales longer than 20 minutes while preserving shorter-term variations caused by flares. Since long-duration flares are typically prominent and unlikely to have been missed in \cite{Bruno_2024}, this method ensures minimal loss of flare signal. A surface variability on timescales shorter than 20 minutes may remain in the light curves, however, and contribute to false positives. 
To optimise the detrending process, we followed \cite{Bruno_2024} and tested polynomial degrees ranging from 2 to 10 in the Savgol filter. For each light curve, we selected the polynomial degree that minimised the Akaike Information Criterion (AIC) in the residuals, ensuring that more active stars, potentially exhibiting stronger modulation from surface activity, were detrended using higher-degree polynomials. Finally, we sigma-clipped outliers at each iteration at a 5$\sigma$ threshold of the flux level in order to minimise the impact of large flares on the detrending process.
We also explored other detrending methods, including non-windowed polynomial fitting, Gaussian processes with a Matérn kernel, and M-estimators using Tukey's biweight function \citep{Hippke_2019}. However, we found these approaches to be less effective at removing time-correlated variations. In particular, Gaussian processes performed the worst, likely due to the short duration of CHEOPS light curves preventing the model from converging. Ultimately, while allowing polynomial orders to vary slightly improved residual scatter, the most significant improvement came from using a windowed detrending approach.
After visual inspection, we removed light curves exhibiting excessively high RMS values ($>$10,000 ppm), likely resulting from a failed photometric extraction and/or detrending, which accounted for $\sim$6\% of the sample. Among the remaining light curves, 84\% were observed at a cadence of 3 seconds, 14\% at 5 to 7 seconds, and the remaining 2\% at cadences longer than 12 seconds. Since the next steps of the analysis involve a denoising process, we refer to the detrended and normalised CHEOPS light curves as 'original' throughout the rest of the paper (as opposed to 'denoised').

\subsection{Flare detection}
\label{subsection:flare_detection}
\begin{figure}[h]
\centering
\includegraphics{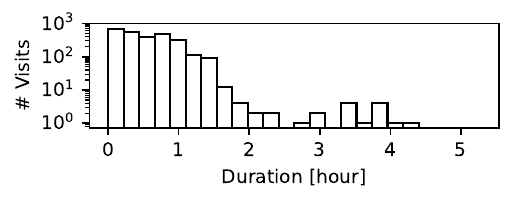}
\caption{Distribution of the visit duration within our sample.}
\label{fig:orbit_durations}
\end{figure}

CHEOPS orbits the Earth with a stable period of 99 minutes, but the available science time per orbit is reduced due to Earth occultations, the South Atlantic Anomaly, and other observational constraints. The uninterrupted science time of an observation is referred to as a visit, and the duration of a visit depends on the target's position in the sky. Targets located near the ecliptic plane can benefit from longer uninterrupted observations, sometimes extending across multiple CHEOPS orbits, while targets near the ecliptic poles usually have shorter visits with frequent interruptions \citep{Benz_2021}. To provide an overview of the actual on-target time composing our sample, Table \ref{tab:stars_distrib} presents the number of stars per spectral subtype, the corresponding number of visits, and the total science time.
A flare candidate cannot be considered as a candidate when it spans two separate visits because data points belonging to the flare profile would be missing from the resulting light curve. Recovering only part of a flare would lead to underestimating the flare energy and biasing the derived flare statistics. Our flare detection algorithm therefore only considered flares whose start and stop times occurred within the same visit. Figure \ref{fig:orbit_durations} shows the distribution of the duration of the visits constituting our sample. Although most of the visits are as expected shorter than 99 minutes, we found a few visits with longer durations of up to 4.5 hours corresponding to targets easily observable by CHEOPS. Since M dwarf flares can have durations of up to $\sim$10 hours \citep{Davenport_2016, Pietras_2022}, conducting flare detection with CHEOPS implies that the population of long-duration flares will not be recovered in this study. 

\begin{table}[h]
\caption{Distribution of stars, visits, and on-target time within our sample according to spectral subtype.}
\centering
\begin{tabular}{cccc}
\hline
Spectral type & \# stars & \# visits & On-target time [d] \\ \hline
M0V           & 33       & 669              & 17.37       \\
M1V           & 27       & 976             & 27.23       \\
M2V           & 17       & 287              & 6.72       \\
M3V           & 13        & 292              & 7.52       \\
M4V           & 15       & 169              & 4.46       \\
M5V           & 4        & 70              & 1.93        \\
M6V           & 1        & 33              & 1.42        \\       
\hline
\end{tabular}
\label{tab:stars_distrib}
\end{table}

The photometric signature of flares is typically characterised by a sharp rise (usually associated with bremsstrahlung radiation) followed by an exponential decay (corresponding to radiative cooling) \citep{Kowalski_2013, Davenport_2014}. An effective way to detect flares is by monitoring the increase in flux they produce in the stellar light curve. 
For this purpose, we employed \texttt{AltaiPony}\footnote{\url{https://altaipony.readthedocs.io/en/latest/}}, a flare detection tool developed for Kepler, K2 and TESS observations\citep{Davenport_2016, Ilin_2022}. It implements a sigma-clipping algorithm based on the flare detection criteria of \cite{Chang_2015}: $N_1$, the minimum positive flux deviation from the median normalised by local scatter; $N_2$, a corresponding minimum deviation accounting for photometric errors; and $N_3$, the minimum number of consecutive data points meeting these conditions. For CHEOPS, we adopted $N_1 = 3$, $N_2 = 2$, and $N_3 = 5$, requiring a minimum of 5 consecutive points above the $3\sigma$ threshold. This setting reduces false positives but excludes flares shorter than 15 seconds.

Since the detection threshold of a sigma-clipping method for flare detection depends on the light curve noise properties, we intend to use a wavelet-based algorithm to denoise our data. This method is expected to lower the effective noise floor while preserving the amplitude and shape of flares, thereby increasing the S/N of low-energy flares and improving their recovery rate. To validate the advantages of this approach, we performed a comparative analysis with the \texttt{PEAKUTILS}\footnote{\url{https://peakutils.readthedocs.io/en/latest/}}-based detection method used in \cite{Bruno_2024}, which combined a peak detection algorithm with a flare profile fitting. In our comparison, we applied both \texttt{AltaiPony} and \texttt{PEAKUTILS}, with and without wavelet-based denoising, using a consistent $3\sigma$ detection threshold. A summary of the results is presented in Table \ref{Altaipony_Peakutils_table}.

\begin{table}[h]
\centering
\caption{Number of flares detected using \texttt{AltaiPony} and \texttt{PEAKUTILS}, with and without wavelet-based denoising.}
\begin{tabular}{|l|l|l|l|}
\hline
          & Original & Denoised & Increase (\%)\\ \hline
\texttt{AltaiPony} & 215      & 291   & +35.3\%   \\ \hline
\texttt{PEAKUTILS} & 229      & 299   & +30.6\%   \\ \hline
\end{tabular}
\label{Altaipony_Peakutils_table}
\end{table}

\begin{figure*}
\centering
\includegraphics[width=0.7\textwidth]{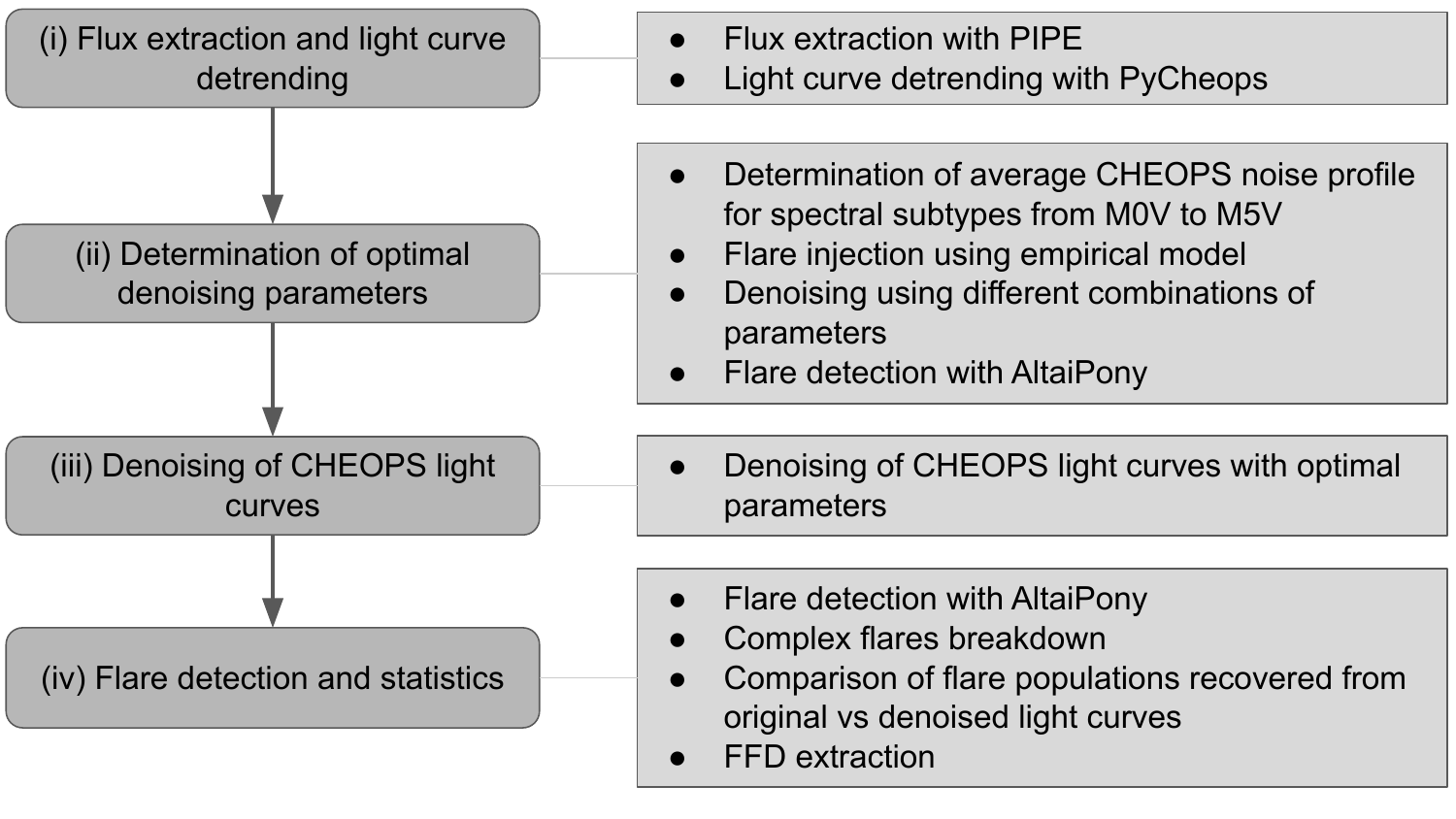}
\caption{Flow diagram of the procedure we followed.}
\label{fig:procedure}
\end{figure*}

While \texttt{PEAKUTILS} detected a slightly higher number of flares overall, it also exhibited a higher false positive rate, especially in the denoised data. Flare injection-recovery tests showed that \texttt{PEAKUTILS} detects weak flares more readily, but also retrieved more spurious detections below $\sim$$10^{27}$ erg. In contrast, \texttt{AltaiPony} offers a more conservative detection by integrating a photometric uncertainty criterion. Its lower false positive rate and sharper detection threshold yield a more homogeneous flare sample. The application of wavelet denoising consistently improved detection rates for both algorithms. However, integrating it within \texttt{AltaiPony} provides a more robust sample for accurately assessing the impact of wavelet-based denoising on flare recovery rates.

Consequently, our specific workflow was to: 
\begin{enumerate}
    \item determine the noise profile of a 'prototypical' M dwarf light curve observed by CHEOPS.
    \item generate a corresponding synthetic light curve in which to inject flares of different energies, durations, and amplitudes. 
    \item determine the denoising parameters that optimise the recovery of the injected flares. 
    \item apply the wavelet-based denoising algorithm with the optimal parameters on the original CHEOPS light curves. 
    \item conduct flare detection on the original and denoised CHEOPS light curves using \texttt{AltaiPony}. 
\end{enumerate}
A flow diagram of the entire procedure we followed is presented in Figure \ref{fig:procedure}.

\subsection{Estimating noise profiles}
\label{subsect:noiseprofile}
\begin{figure*}
\centering
\includegraphics[width=0.8\textwidth]{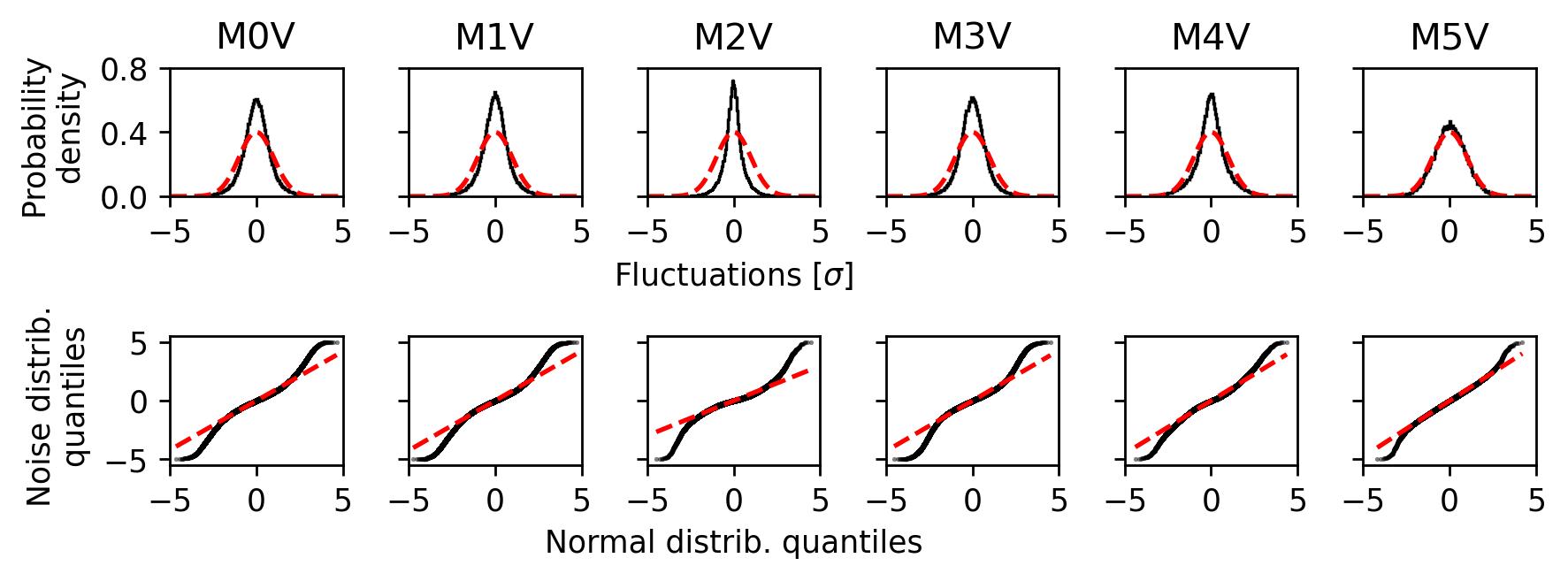}
\caption{Averaged noise representation for each spectral type. In the top row, the dashed red curves correspond to a normal distribution, and the black histograms correspond to the noise distribution histograms in the CHEOPS light curves. In the bottom row, the dashed red lines correspond to the quantiles of a normal distribution, and the black curves correspond to the quantiles of the noise distribution in the CHEOPS light curves.}
\label{fig:noises}
\end{figure*}

To determine the typical noise profile of a CHEOPS light curve, we computed the average noise profile for each spectral subtype by averaging the photometric fluctuations from the median value of each light curve in our sample, following the method of \cite{Jess_2019, Dillon_2020} and \cite{Grant_2023}. These profiles therefore quantify the averaged white noise and time-correlated scatter at timescales shorter than 20 minutes for each spectral subtype. We discarded the M6V subtype since only one star was included in this group (see Table \ref{tab:stars_distrib}). We therefore included targets of spectral type ranging from M0V to M5V. Figure \ref{fig:noises} displays the averaged fluctuations for each spectral subtype and compare them against a normal distribution. 
The resulting noise distributions for each spectral subtype appear thinner and taller than a normal distribution, with more pronounced tails, indicating a leptokurtic nature. In statistics, the kurtosis is a measure of the shape of a probability distribution, specifically how the tails and peak compare to a normal distribution. A leptokurtic distribution (one with positive kurtosis) has a sharper peak and heavier tails than a normal distribution, indicating more extreme deviations from the mean. This is confirmed by the quantile-quantile (Q-Q) plots for each spectral subtype in the second row of Figure \ref{fig:noises}, which compare the observed data distribution to a theoretical normal distribution. If the data followed a normal distribution, the points would align along the diagonal. However, the deviation in the tails indicates an excess of outliers (see Figure 3 of~\cite{Jess_2019} for a collection of different distribution types). A leptokurtic noise distribution has several implications for the light curves and the flare detection. First, the presence of heavier tails means that extreme fluctuations occur more frequently than expected under a normal distribution. This could mimic or obscure low-amplitude flare events, complicating their detection. The shapes of the noise distributions indicate that the photometric noise in the light curves is not purely Gaussian and that the remaining time-correlated scatter contributes to a higher frequency of outliers.
Consequently, the denoising process must accurately account for these leptokurtic features to efficiently distinguish between data points caused by flares and those arising from noise artefacts. Table \ref{Table:noise_parameters} compiles the mean value, standard deviation, skewness, and Fisher kurtosis of the averaged noise distribution for spectral subtype. Understanding these specific noise characteristics allows us to refine the denoising and flare detection algorithms, reducing false positives and improving flare detection reliability.

\begin{figure*}
\centering
\includegraphics[width=0.8\textwidth]{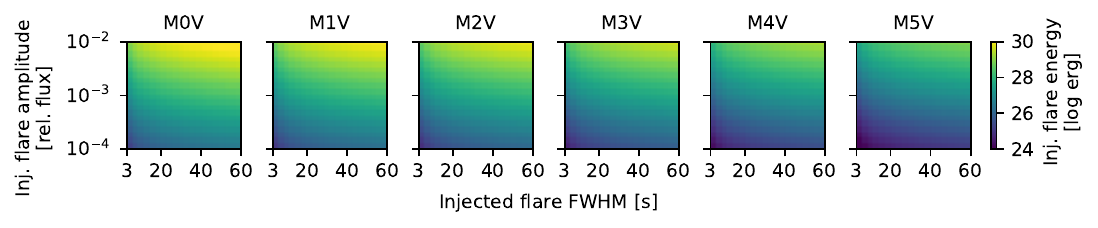}
\caption{Injected flare energy matrix as a function of the injected flare amplitude, FWHM, and spectral subtype.}
\label{fig:flare_energies}
\end{figure*}

\begin{table}
\caption{Mean, standard deviation, skewness, and Fisher kurtosis values of the noise profile of an average CHEOPS light curve corresponding to each spectral subtype.}
\begin{tabular}{ccccc}
\hline
Spectral type & Mean  & St. dev & Skewness & Kurtosis \\ \hline
M0V           & 0.000 & 0.002   & 0.001    & 1.090    \\
M1V           & 0.000 & 0.001   & 0.009    & 1.294    \\
M2V           & 0.000 & 0.002   & 0.063    & 2.528    \\
M3V           & 0.000 & 0.001   & 0.021    & 0.959    \\
M4V           & 0.000 & 0.002   & 0.029    & 1.233    \\
M5V           & 0.000 & 0.002   & 0.010    & 0.038    \\ \hline
\end{tabular}
\label{Table:noise_parameters}
\end{table}

\subsection{Injecting synthetic flares}
\label{subsection:injection_and_recovery}
To create synthetic light curves for each spectral subtype, we first injected into empty light curves the noise distribution corresponding to each spectral subtype using the parameters compiled in Table \ref{Table:noise_parameters}. We then injected flares of known energies, amplitudes, and durations in the light curves corresponding to each spectral subtype. For this, we started by determining the stellar parameters of a 'prototypical star' from M0V to M5V. The effective temperatures were determined by calculating the average temperature of the stars in our sample within each spectral subtype. The radius values for each spectral subtype were sourced from \cite{Pecaut_2013}. The quiescent luminosities were calculated using the Stefan-Boltzmann law with the obtained effective temperatures and radii. We compile the average effective temperature, radius, and luminosity for each spectral subtype in Table \ref{Table:average_stellar_params}. 

\begin{table}
\caption{Average effective temperature, radius, and luminosity for each spectral type.}
\centering
\begin{tabular}{cccc}
\hline
Spectral type & $T_{\mathrm{eff}}$ {[}K{]} & Radius {[{$R_{\odot}$}]} & Luminosity {[{$L_{\odot}$}]}\\ \hline
M0V           & 3850                & 0.588  & 0.068      \\
M1V           & 3680                & 0.501  & 0.041      \\
M2V           & 3550                & 0.446  & 0.028      \\
M3V           & 3400                & 0.361  & 0.016      \\
M4V           & 3200                & 0.274  & 0.007      \\
M5V           & 3050                & 0.196  & 0.003      \\ \hline
\end{tabular}
\label{Table:average_stellar_params}
\end{table}

Flares were simulated using the package \texttt{Llamaradas Estelares}\footnote{\url{https://github.com/lupitatovar/Llamaradas-Estelares}} \citep{Mendoza_2022}. This package simulates the morphology of white-light flares with a template derived from a set of flares observed in GJ 1243 based on two parameters: relative amplitude and FWHM (Full Width at Half Maximum, serving as a proxy for flare duration). We could simulate flares with different energies by using a known range of parameters for the amplitude and the FWHM. Following \cite{Davenport_2016}, flare energies were calculated by multiplying the Equivalent Duration (ED) by the stellar quiescent luminosity. \cite{Bruno_2024} found that the sensitivity of CHEOPS could detect flares with energies down to $\sim$$10^{27}$ erg. We determined that relative flare amplitudes ranging between [$10^{-4} - 10^{-2}$] and FWHM between [0 - 60] seconds could simulate flares of energies within [$10^{24} - 10^{30}$] erg, depending on the quiescent luminosity of the star. Simulating flares within this energy range would enable us to test the impact of denoising on the recovery rate around this detection threshold. Figure \ref{fig:flare_energies} displays the injected flare energy as a function of the flare amplitude, FWHM, and spectral subtype of the star. 

\subsection{Determining the optimal denoising parameters}
\label{subsection:denoising}

\begin{figure*}
\centering
\includegraphics[width=0.77\textwidth]{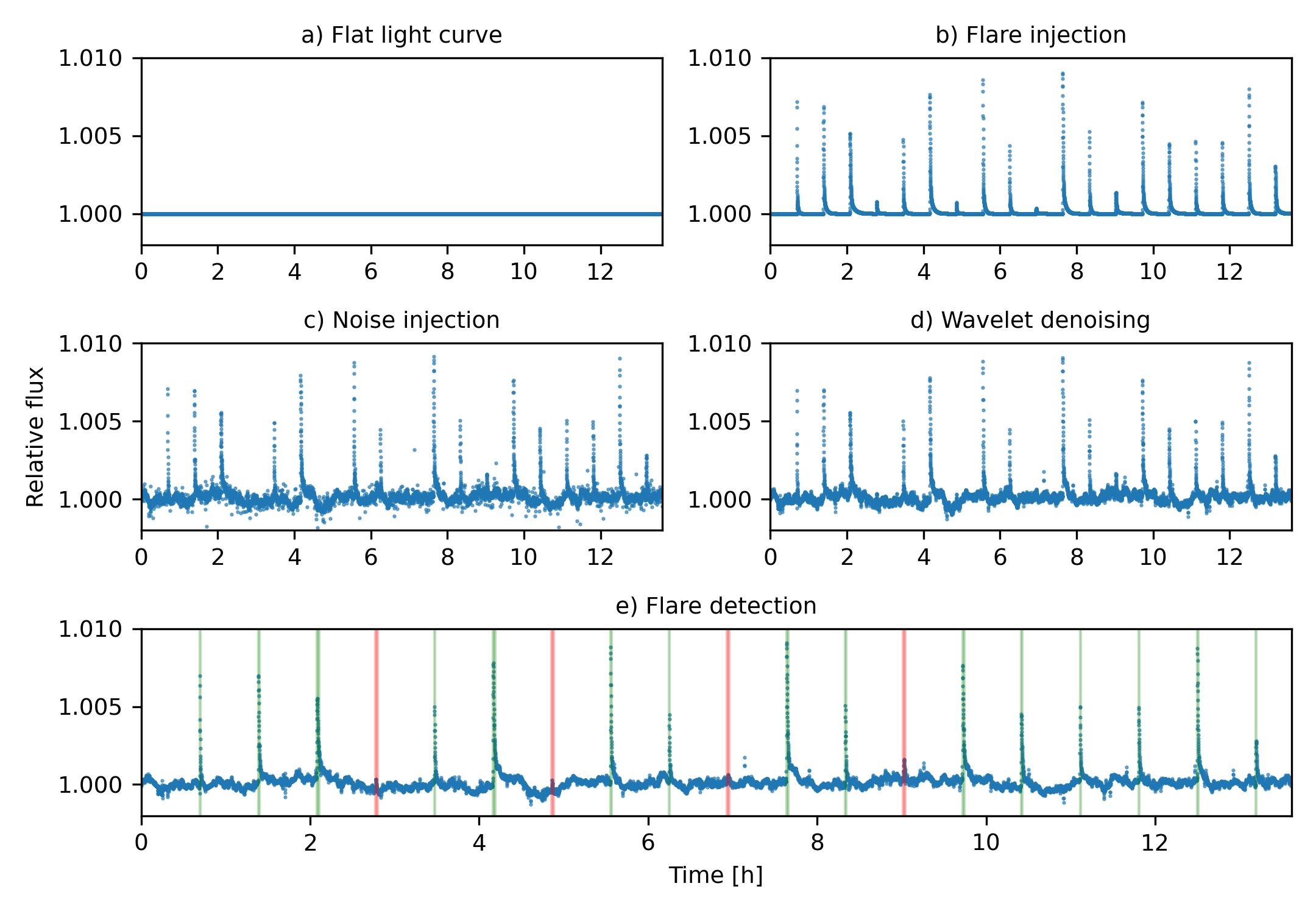}
\caption{Step-by-step representation of the flare injection, noise injection, wavelet denoising, and flare recovery process. Only 20 flares are shown to better visualise the flux variations in the light curve. In this example, the wavelet-based denoising was conducted with the Haar wavelet and a decomposition level of 2. Only 16 out of 19 flares are recovered in this portion of the light curve. The recovered and lost flare candidates are highlighted with a green and red background, respectively.}
\label{fig:injections}
\end{figure*}

The Discrete Wavelet Transform (DWT) is commonly used to decompose and denoise time-series data. This is done by scaling and translating a mother wavelet across the data, resulting in a decomposition into different frequency bands through low-pass and high-pass filtering, followed by a downsampling. Stochastic noise, typically in high-frequency components, is thresholded and removed before reconstructing the data by using the inverse transform. Unlike other denoising methods based on the Fourier Transform, the DWT retains both temporal and frequency information, which is beneficial for recovering non-stationary signals \citep{Mallat_2009}. However, the Stationary Wavelet Transform (SWT) is often preferred for denoising as it avoids downsampling, maintaining the signal length at each decomposition level through an \textit{à trous} algorithm. This approach offers a redundant, shift-invariant representation that enhances noise reduction accuracy and effectively preserves signal features, resulting in more consistent and reliable denoising results \citep{Percival_2000, Fors_2006}. 

The threshold value in wavelet-based denoising is key to distinguish between noise and signal by determining which wavelet coefficients to suppress or retain. A well-chosen threshold effectively reduces noise while preserving essential signal features. However, an overly high threshold can erase important details, while a low threshold may fail to remove enough noise.
Determining the threshold value typically involves statistical methods, such as the universal threshold proposed by \cite{Dohono_1994}. Its logarithmic scaling with data length ensures effective noise filtering for larger time-series while preserving signal features \citep{Abramovich_1995, Percival_2000}. We adapted this universal threshold by replacing its dependence on the standard deviation of the signal by the median absolute deviation, which is more robust to outliers. This adjustment effectively balances noise reduction and signal preservation, especially in the presence of non-Gaussian noise \citep{DelSer_2018}. 

The decomposition level determines the number of times the data is iteratively decomposed into wavelet components, with higher levels providing finer resolution of the signal and stronger denoising. Each decomposition level corresponds to different frequency scales in the data, enabling multi-resolution analysis that typically separates noise from significant signal features. The decomposition level is therefore generally chosen based on the number of data points in the input data. An appropriate decomposition level allows us to effectively separate noise from the signal while preserving key signal characteristics. A level that is too low may fail to remove enough noise, while a level that is too high may suppress important signal details and introduce artefacts from the mother wavelet function \citep{Yang_2016}.
Since each decomposition level halves the data frequency, consequently capturing more detail at lower frequencies, wavelet decomposition follows a binary scaling \citep{Mallat_1989}. The maximum number of useful decomposition levels $J$ therefore scales logarithmically with the number of data points $N$, such that $2^J \leq N$. This ensures effective noise isolation while preserving the signal integrity \citep{Daubechies_1992}. 
CHEOPS visits vary in the number of data points based on visit duration and observing cadence. The decomposition level is tailored to the number of data points in each visit, ensuring shorter visits are decomposed at a lower level than longer ones. This optimisation improves the denoising process and ensures effective use of the SWT across the different visit durations in our sample.

\begin{figure*}
\centering
\includegraphics[width=0.875\textwidth]{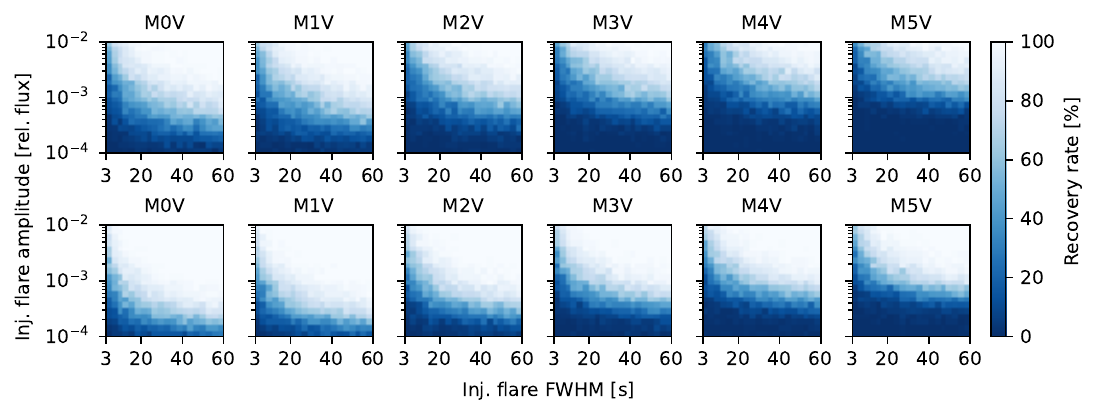}
\caption{Flare recovery rate as a function of the injected flare amplitude, FWHM, and spectral subtype after denoising with the Haar mother wavelet and a decomposition level of 2. The top row corresponds to the recovery rates before denoising, and the bottom row shows the recovery rates after denoising.}
\label{fig:recovery}
\end{figure*}

\begin{figure*}
\centering
\includegraphics[width=0.875\textwidth]{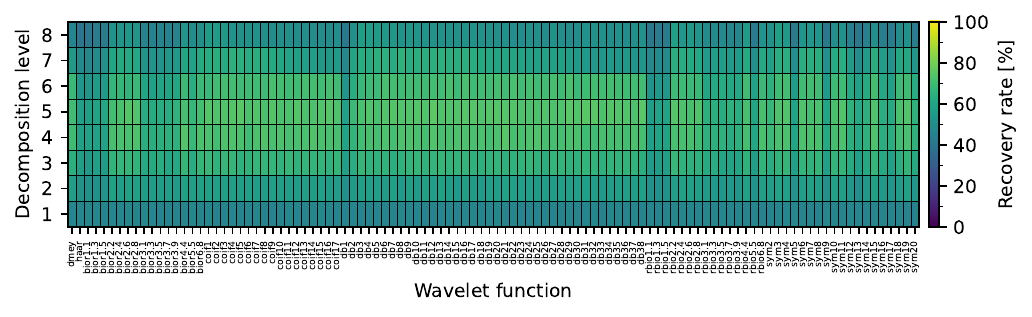}
\caption{Flare recovery rate, averaged across all spectral subtypes, as a function of the selected mother wavelet and decomposition level.}
\label{fig:heatmap}
\end{figure*}

The choice of the mother wavelet is crucial in wavelet-based denoising, as it affects the accuracy of noise separation from the signal. Wavelets with short support, such as the \textit{Haar} wavelet, are efficient at detecting sharp discontinuities but may introduce high-frequency noise, whereas wavelets with longer supports, like the \textit{Daubechies} and \textit{Symlets}, are better at capturing smooth variations but may blur sharp features \citep{Haar_1910, Mallat_1989, Daubechies_1992}. Vanishing moments refer to the ability of a wavelet to represent polynomials up to a certain degree. Wavelets with higher vanishing moments are more effective at filtering noise but are computationally demanding and may lead to overfitting \citep{Dohono_1994}. For the SWT, discrete and biorthogonal wavelets are used to ensure accurate data reconstruction, non-redundancy, and minimal introduction of artefacts \citep{Chui_1992}.
To identify the optimal parameters for denoising CHEOPS light curves, we followed a systematic approach, illustrated in Figure \ref{fig:injections}:
\begin{enumerate}
    \item We created synthetic light curves for each spectral subtype and injected flares with energies from $10^{24}$ to $10^{30}$ erg, simulating various flare profiles. For each energy bin, spanning $10^{25}$ erg, we simulated 1,000 flares using the amplitude-FWHM parameters described in Figure \ref{fig:flare_energies}. We separated flare peaks by 45 minutes to ensure that the flux had returned to a quiescent level before the beginning of the next flare.
    \item We injected a noise profile corresponding to each spectral subtype according to the distribution parameters compiled in Table \ref{Table:noise_parameters}.
    \item We conducted a SWT-based denoising using the previously mentioned threshold for each discrete and biorthogonal mother wavelet included in the package \texttt{PyWavelets} \citep{Lee_2019}. This included wavelets from the \textit{Haar} (haar), \textit{Daubechies} (db), \textit{Symlets} (sym), \textit{Coiflets} (coif), and \textit{Biorthogonal} (bior) families, with decomposition levels from 1 to 8. Detailed properties of all the considered wavelets are compiled in Table~\ref{wavelet_list}.
    \item We detected flares in the resulting light curves using the flare detection algorithm described in Section \ref{subsection:flare_detection}.
\end{enumerate}

As an example of the output of this process, Figure \ref{fig:recovery} presents the improvement in flare recovery rate across all spectral subtypes after denoising with the Haar wavelet at a decomposition level of 2. The primary limitation for flare recovery is the amplitude, with a sharp decline in recovery rates for very low amplitude flares. This behaviour is expected, as even after denoising, low amplitude flares are unlikely to exceed the $3\sigma$ detection threshold above the noise floor and fulfil the $N_1$ criterion. In addition, recovery rates also drop for very low FWHM, likely due to the flares not satisfying the $N_3$ criterion due to insufficient consecutive data points above the detection threshold. This suggests that even though flares of different impulses can have the same energy (see Figure \ref{fig:flare_energies}), high-impulse flares generally have a higher recovery rate compared to low-impulse ones, provided they have durations longer than $N_3$ data points. Overall, the denoising process improves recovery rates across all spectral subtypes, enabling the detection of flares with amplitudes smaller by approximately half an order of magnitude than those identified in the original light curves. 

\begin{figure}
\centering
\includegraphics[width=0.45\textwidth]{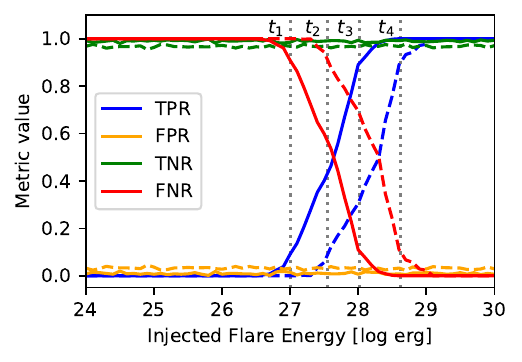}
\caption{Evolution of confusion matrix metrics as a function of the injected flare energy. The TPR is shown in blue, the FPR in orange, the TNR in green, and the FNR in red. Metrics for the original light curves are represented by dashed lines, while those for the denoised light curves are shown as solid lines. The vertical black dotted lines indicate the four detection thresholds.}
\label{fig:confusion_matrices_evolution2}
\end{figure}

To compare the gain in recovery rate obtained after denoising with different sets of parameters, we averaged the recovery rate across spectral type for each wavelet and decomposition level. Figure \ref{fig:heatmap} displays the average flare recovery rate obtained for all flares between $10^{24}$ and $10^{30}$ erg after denoising with different wavelets and decomposition levels. Recovery rates corresponding to each spectral subtype can be found in Figure~\ref{annex:recovery_rates}. Although a single wavelet did not produce a significant improvement in recovery rate compared to others, \textit{Coiflets} and \textit{Daubechies} wavelets consistently provided the highest recovery rate across several vanishing moments. Given their frequent use in analysing astrophysical datasets, we selected \textit{Daubechies} wavelets for denoising the CHEOPS light curves \citep{Ojeda_2014, Souza_2018, Bolzan_2020}. Specifically, we selected the db18, which provides a balance between a high and low number of vanishing moments. 

\begin{figure}
\centering
\includegraphics[width=0.75\columnwidth]{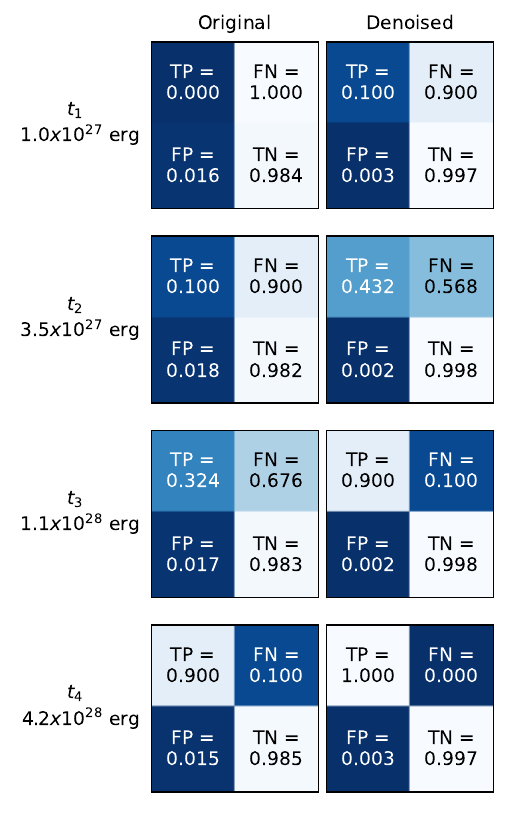}
\caption{Confusion matrices obtained from the flare injection and recovery process for the original (left) and denoised (right) light curves. The matrices are shown at the four detection thresholds, corresponding to TPR rises above 10\% ($t_1$ and $t_2$) and reaching 90\% ($t_3$ and $t_4$), as defined in the text.}
\label{fig:confusion_matrices_tresholds}
\end{figure}

An important parameter of the SWT is the coefficient thresholding method. Hard thresholding sets coefficients below the threshold to zero, preserving signal features but potentially introducing discontinuities and artefacts in the reconstructed signal. Soft thresholding, on the other hand, not only sets coefficients below the threshold to zero but also shrinks the remaining coefficients towards zero, resulting in smoother data with fewer artefacts. However, it can slightly attenuate preserved features. Generally, hard thresholding is better for retaining sharp features, while soft thresholding improves signal smoothness and continuity. 
We compared the impact of both thresholding methods on the recovery rate and found only a marginal difference of less than 1\% within each energy bin. We therefore chose to use a hard thresholding to denoise the CHEOPS light curves for several reasons. The signal of low-amplitude flares could be affected by the shrinkage introduced by soft thresholding, which would lead to the non-detection of the flares reaching just above the detection threshold. Additionally, the shrinkage could potentially lead to inaccurate energy estimations of the detected flares. Finally, hard thresholding is less computationally intensive than soft thresholding because it requires one operation less on each coefficient value. This makes it a more efficient method when it is applied to large datasets. 

Our flare detection algorithm essentially functions as a binary classifier, distinguishing between 'flaring' and 'non-flaring' states. To evaluate its performance, we generated 1000 flare-free light curves for each $10^{25}$ erg energy bin and applied the detection algorithm to measure its classification efficiency. Specifically, we constructed confusion matrices for each injected flare energy bin, defining four possible outcomes: true positive (TP), where a flare was injected and correctly retrieved; false positive (FP), where no flare was injected but one was erroneously retrieved; false negative (FN), where a flare was injected but not retrieved; and true negative (TN), where no flare was injected and none was retrieved. We then calculated the true positive rate (TPR), false positive rate (FPR), false negative rate (FNR), and true negative rate (TNR) in function of the injected flare energy. 

In Figure \ref{fig:confusion_matrices_evolution2}, we present the evolution of these rates, averaged across all spectral types, for both the original and denoised light curves. We observed that the false positive rate (FPR) remained independent of the injected flare energy but instead depended on the noise level in the light curves. Notably, the FPR remained consistently low across the entire energy range: below 0.025 for non-denoised light curves and below 0.010 for denoised light curves. 

The denoising process significantly enhanced the true positive rate (TPR), effectively shifting the detection threshold towards lower injected flare energies by nearly half an order of magnitude. To quantify these improvements, we identified four key thresholds: 
\begin{itemize}
    \item $t_1$: where the TPR for denoised light curves exceeds 10\% ($1.0\times10^{27}$ erg),
    \item $t_2$: where the TPR for original light curves exceeds 10\% ($3.5\times10^{27}$ erg),
    \item $t_3$: where the TPR for denoised light curves reaches 90\% ($1.1\times10^{28}$ erg),
    \item $t_4$: where the TPR for original light curves reaches 90\% ($4.2\times10^{28}$ erg).
\end{itemize}
Figure \ref{fig:confusion_matrices_tresholds} illustrates the confusion matrices at these thresholds. Consistent with the trend shown in Figure \ref{fig:confusion_matrices_evolution2}, the denoising process significantly improves the TPR between $t_1$ and $t_4$, enabling a more complete detection of flares at lower energies compared to the original light curves.

\subsection{Flare detection in denoised light curves}
We then denoised all CHEOPS light curves (further referred to as 'denoised'), separated by visit and performed the flare detection described in Section \ref{subsection:flare_detection}. Following the method of \citet{Raetz_2020,Bruno_2024} and \citet{Fortier_2024}, we determined the quiescent luminosity of each star in the sample by adopting the CHEOPS zero-point and effective wavelengths available on the filter profile service of the Spanish Virtual Observatory \citep{Rodrigo_2012} as follows: 
\begin{equation}
    L_0 = 4 \pi d^2 \times Q_{cheops} (F_{vega} \times 10^{-0.4 \times G})
\end{equation}
where $d$ is the distance of the star, $Q_{cheops}$ is the quantum efficiency of the CHEOPS sensor integrated over the 330-1100nm wavelength range, $F_{vega}$ is the CHEOPS flux of Vega (CHEOPS zero-point), and $G$ is the Gaia $G$-band magnitude of the star. We also derived several flare parameters, such as the flare duration and FWHM, flare impulse (amplitude / FWHM), and flare peak luminosity (amplitude $\times$ quiescent luminosity).

\subsection{Complex flare breakdown}
To break down complex flares displaying several visible peaks, we consider that such events can be described as a superposition of several simple flares, which can be modelled by the empirical flare template of \texttt{Llamaradas Estelares}. For a given complex flare event, the objective is therefore to automatically determine the number of flares and their parameters (namely $t_{peak}$, amplitude, and FWHM) that best describe it. For this, we developed an algorithm to iteratively fit up to 5 flares for every flaring event detected using \texttt{AltaiPony}. The flare region to fit for each event included 15 seconds before the start time and 45 seconds after the stop time of the flare to ensure the flux fully returned to the quiescent level within the fitting window. Following the procedure of \cite{Davenport_2014}, we imposed a first guess and boundaries for the fitted flare parameters. We seeded the parameters of the fitting by using the peak flux amplitude and time and 15\% of the flare full duration for the FWHM. The relative amplitude of each component was required to be larger than $10^{-4}$, being the flare amplitude after which the recovery rate dropped to 0\% across all spectral types (see Figure \ref{fig:recovery}). We also required $t_{peak}$ to occur within the boundaries of the flare window, and FWHM to be larger than the observing cadence of the light curve and smaller than 50\% of the flare total duration. No priors on the relations between flare amplitude, FWHM, and $t_{peak}$ were included. Following \cite{Bruno_2024}, we used the AIC to determine the best fit, which took the form 
\begin{equation}
    AIC_n = M \times ln(\frac{\chi^2}{M}) + 2k_n
\end{equation}
where M is the number of data points that fell within the flare window, and $k_n$ is the number of degrees of freedom of the $n$-th model. This statistic determines the improvement of the fit (decreased $\chi^2$) while penalising the increasing number of free parameters used in the model. 
We chose to use the AIC instead of the BIC like \cite{Davenport_2014} because the AIC penalises less the number of components. The AIC therefore tends to favour more complex models with a higher number of parameters, which would enable further breakdown. Finally, the best solution fit was selected as the $n$-th model with the smallest AIC to have decreased by at least 6 units from the previous $(n - 1)$th model. The choice of a 6-unit improvement threshold was determined in \cite{Bruno_2024} by visual inspection of complex flare events to ensure the algorithm did not overfit them. 

\section{Results}
\label{section:results}

We began our analysis by applying the flare detection algorithm to the detrended CHEOPS light curves, initially identifying 215 flares with energies ranging from $4.2\times10^{26}$ erg to $8.9\times10^{30}$ erg. To optimise the detection process, we employed a SWT-based denoising algorithm using the db18 wavelet base, as detailed in Section \ref{subsection:denoising}, to reduce noise within each CHEOPS visit. We then reran the flare detection algorithm on the denoised light curves and detected a total of 291 flares with energies ranging between $3.7\,\times\,10^{26}$ erg and $8.9\,\times\,10^{30}$ erg. After running the flare-breakdown algorithm, 101 individual components were recovered from 48 complex events in the original light curves, with energies ranging from $5.8\,\times\,10^{26}$ erg to $4.1\,\times\,10^{30}$ erg. 271 individual flare components were recovered in the denoised light curves from 123 complex events, with energies ranging from $1.9\times10^{26}$ erg to $4.1 \times 10^{30}$ erg. The denoising process resulted in a significant increase in the overall flare recovery of $\sim$35\% and improved the recovery of individual flare components by $\sim$64\%. For complex flares, the number of recovered components increased by $\sim$168\%. Eight events (less than 2\% of the denoised flare sample) had recovered energies lower than $10^{27}$ erg, therefore belonging to the upper end of the micro-flare energy range.

\begin{figure*}
\centering
\begin{subfigure}{\textwidth}
\centering
\includegraphics{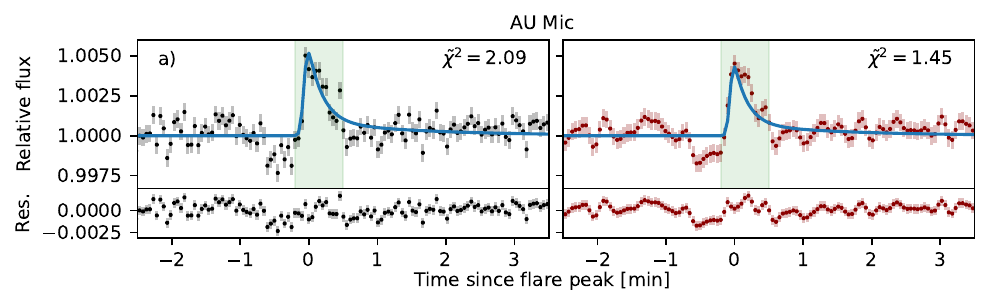}
\end{subfigure}

\begin{subfigure}{\textwidth}
\centering
\includegraphics{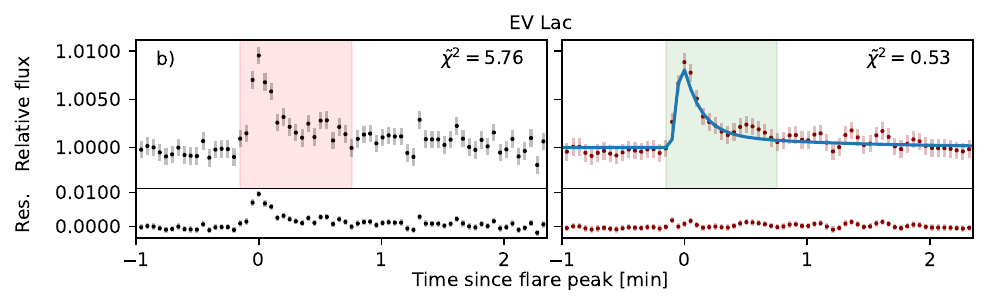}
\end{subfigure}

\begin{subfigure}{\textwidth}
\centering
\includegraphics{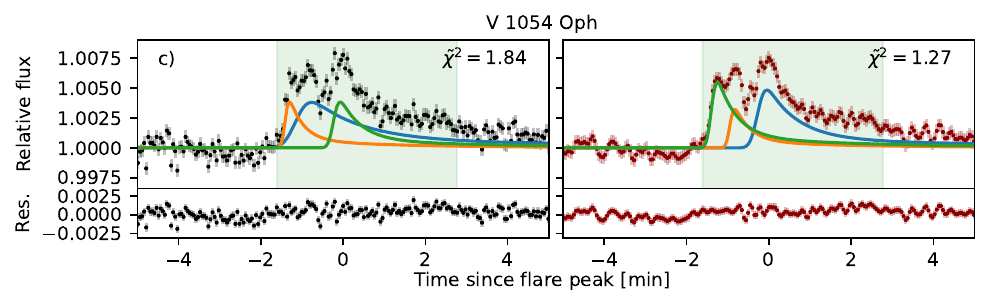}
\end{subfigure}

\begin{subfigure}{\textwidth}
\centering
\includegraphics{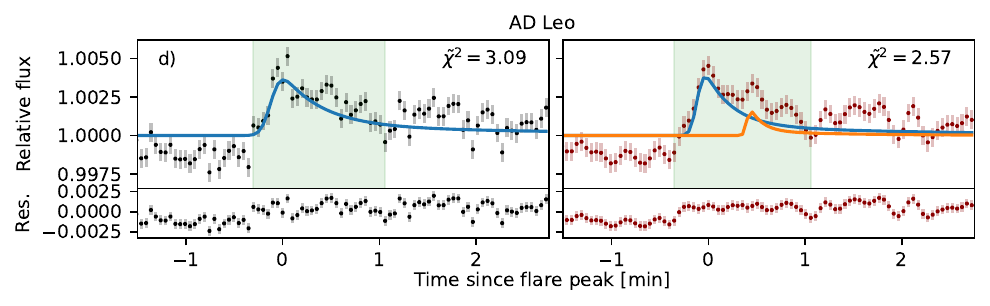}
\end{subfigure}
\caption{Comparison of the flare recovery before (left) and after (right) denoising. In each panel, the original light curve (black data points) is compared to the denoised light curve (red data points). The vertical bars indicate the photometric uncertainty associated with each flux measurement, as estimated during the photometry extraction. The star on which each event was detected is mentioned in the corresponding panel title. The green shaded areas correspond to the data points flagged as belonging to a flare candidate by the flare detection algorithm, and the red shaded areas correspond to a non-detection in the same portion of the light curve. The solid lines represent the fitted flare components. Below each light curve, we show the residuals after subtracting the best-fit flare model. In the top right corner of each panel, the $\tilde{\chi}^2$ calculated on the portion of the light curve we show is given as indication. Each panel group corresponds to a specific case: a) Simple flare detected in the original and denoised light curves. b) Simple flare detected only after denoising. c) Complex flare with all components recovered in the original and denoised light curves. d) Complex flare with an additional component recovered after denoising.}
\label{fig:grid}
\end{figure*}

In Figure \ref{fig:grid}, we present examples of the improved flare recovery enabled by wavelet-based denoising. As shown in panel b), the denoising allowed the recovery of a low-amplitude flare from EV Lac by smoothing out the local scatter caused by the light curve noise, which had previously prevented the flare from meeting the $N_3$ criterion. Flares slightly below the detection limit therefore become detectable as they fulfil the $N_{1,2,3}$ criteria. As shown in panel d), the denoising also allowed for the identification of an additional flare component in a complex flare from AD Leo, which had been merged into a single one due to the light curve noise. This increased sensitivity might allow us to recover a larger population of complex flares, with much lower amplitudes and durations than those detected before applying the denoising.

We also observed several events with well-known flare morphologies. Panel a) of Figure~\ref{fig:grid} shows a pre-flare dip from AU Mic, with an amplitude comparable to the one of the subsequent flare, while panel c) presents a QPP candidate with a sub-minute period from V1054 Oph. Such pre-flare dips and QPPs in flaring light curves have been extensively discussed in~\cite{Bruno_2024}. We present more pre-flare dip detections and QPP candidates from GJ 317, GJ 3323, AU Mic, and AD Leo in Figure~\ref{annex:dip+qpp}. Furthermore, we detected many flare candidates with the typical 'peak-bump' morphology, which were broken down by our algorithm into two distinct components. While some of these events are explicable as a random superposition of two sympathetic flares, \cite{Tovmassian_2003} claims that most peak-bump events are caused by a single flare described by a two-phase underlying emission mechanism, as energy is re-radiated by the stellar photosphere after the peak phase. The complete profile should only be observable when the emission site is close to the centre of the visible stellar disk, however. We also found a few occurrences where the 'bump' occurs before the 'peak', however, which seems to contradict this hypothesis. Finally, a few cases of 'flat-top' structures were also found, although this could be due to a typical 'peak-bump' shape with a lower contrast between the 'peak' and the 'bump' phases. Examples of such flare morphologies from Gl 841 A, V1054 Oph, and Ross 733 are provided in Figure~\ref{annex:morphologies}.  

\begin{figure}
\centering
\includegraphics{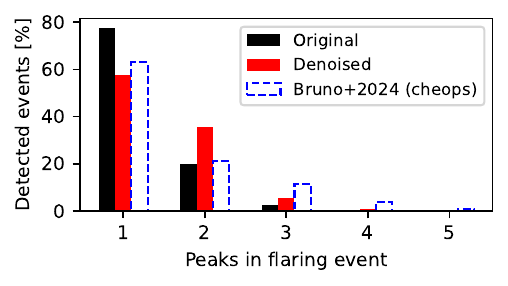}
\caption{Percentages of peaks observed per flaring event. The black and red bars correspond to flares recovered in original and denoised light curves, respectively. We compare our obtained percentages to the ones of \cite{Bruno_2024}, displayed as dashed blue bars.}
\label{fig:peaks}
\end{figure}

\begin{figure}
\centering
\includegraphics{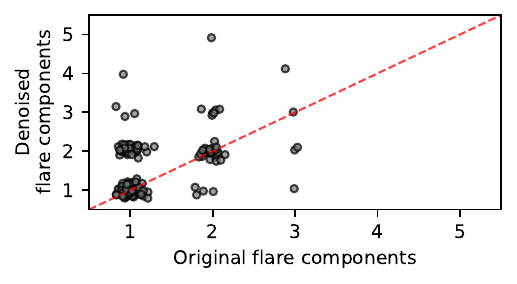}
\caption{Comparison of the number of peaks detected in each flare before and after denoising. Each point corresponds to a single flare. The red dashed line indicates the 1:1 relation, where the number of peaks remains unchanged. Points above the line represent flares that gained components after denoising, while those below lost components.}
\label{fig:morphology}
\end{figure}

\begin{figure}
\centering
\includegraphics{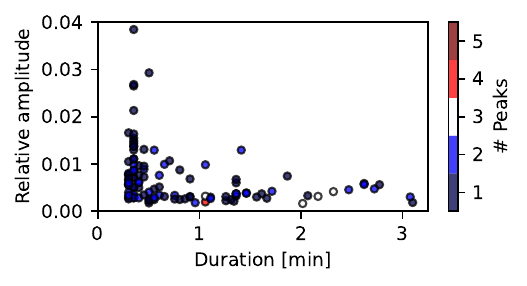}
\caption{Amplitude vs duration of newly detected flares identified only after denoising. Each point represents a flare, with colours indicating the number of components retrieved from the flare breakdown.}
\label{fig:new_flares}
\end{figure}

Figure \ref{fig:peaks} displays the number of peaks observed per flaring event. We find that $\sim$42\% of flares have complex structures composed of multiple peaks, in agreement with previous studies on M stars \citep{Davenport_2014, Howard_2022, Bruno_2024}. In Figure \ref{fig:morphology}, we show the number of components detected in flares that were observed both before and after denoising. In most cases, denoising allows us to identify additional components and enables us to break down complex flare structures in more detail. We compare the percentage of peaks per flaring event with those obtained by \cite{Bruno_2024}, who conducted a similar analysis on CHEOPS light curves from the same programmes. We find that our method retrieved less 3- to 5-peak flares than \cite{Bruno_2024}, which we attribute to two factors. First, a minor difference in determining the 'best-fit' within the breakdown algorithm: we required each $(n-1)$th model to improve the AIC by at least 6 units before exploring the $n$-th model, similar to the approach used by \cite{Davenport_2014} to prevent overfitting. Second, as shown in Figure \ref{fig:new_flares}, denoising facilitates the detection of new flares that were not identified in the original sample. These newly detected flares are generally of low amplitude and short duration, making them more challenging to decompose into multiple components, and contributing to a higher proportion of single- and two-peak flares in the denoised sample.

\begin{figure*}
\centering
\includegraphics{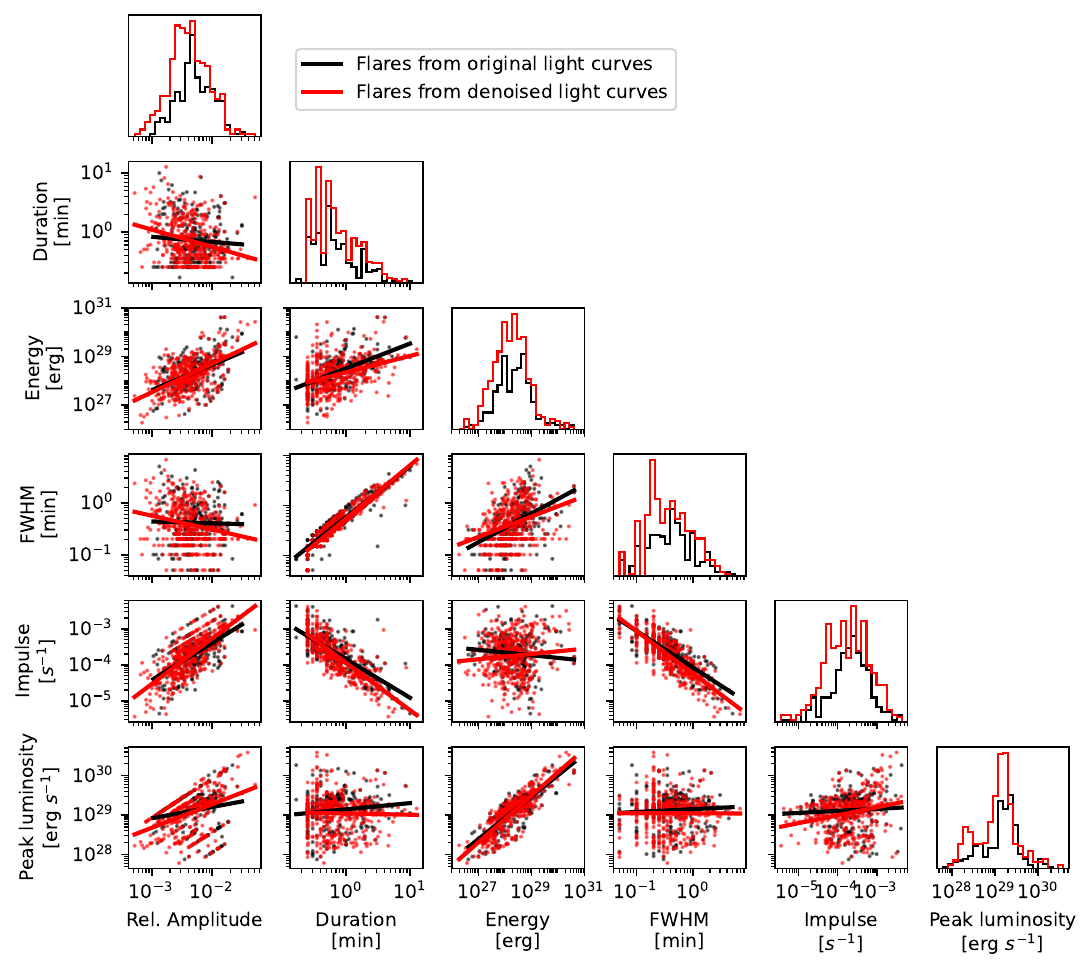}
\caption{Corner plot of the different flare properties, namely relative amplitude, duration, measured energy, FWHM, impulse, and peak luminosity. The black data points represent metrics of individual flare components detected before denoising, while red data points correspond to those detected after denoising. A linear fit is provided for reference in each distribution. The discrete bins for flare duration and FWHM are due to the observational cadence equal to 3 seconds.}
\label{fig:corner}
\end{figure*}

\begin{figure}
\centering
\includegraphics{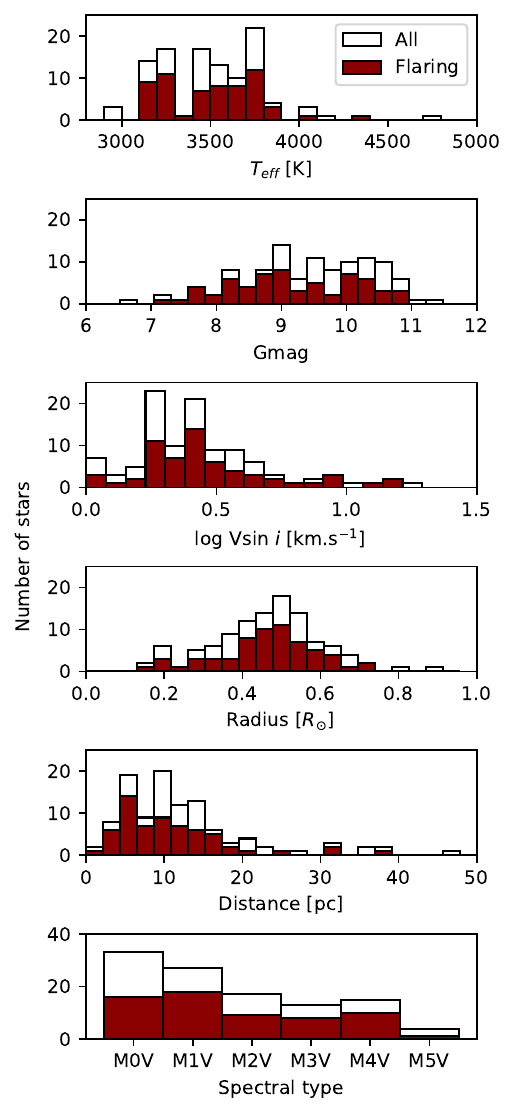}
\caption{Comparison of Gmag, $T_{eff}$, radius, log $v$sin$i$, distance and spectral type for flaring stars (red bars) against the entire stellar sample (white bars).}
\label{fig:flaring_stars_parameters}
\end{figure}

\begin{figure}
\centering
\includegraphics{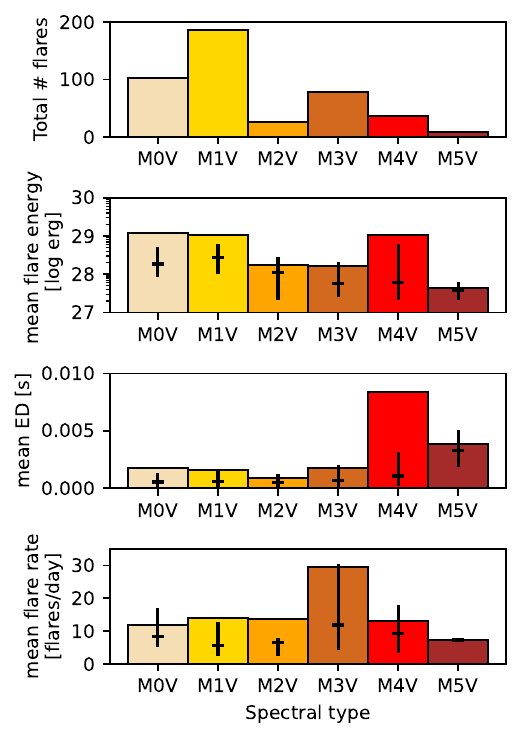}
\caption{Flare statistics by spectral subtype for the stars in our sample. The top panel shows the mean number of flares per star, the middle panel displays the mean flare energy, and the bottom panel presents the mean flare equivalent duration. Vertical bars represent the interquartile range (1st to 3rd quartiles), while horizontal bars indicate the median values. In the M5V bin of the bottom panel, only one star is included, resulting in equal values for the 1st quartile, median, and 3rd quartile.}
\label{fig:flaring_stars_parameters_per_sptype}
\end{figure}

In Figure \ref{fig:corner}, we present a corner plot of the parameters of the recovered flares before and after denoising. In both cases, the energy of recovered flares spans from $\sim$$10^{26}$ erg to $\sim$$10^{31}$ erg. Most flares exhibit relative amplitudes between $10^{-3}$ and $10^{-1}$. Flare durations range between 10 seconds and 20 minutes. Most flares display impulses between $10^{-5}$ and $10^{-3}$ s$^{-1}$. The distributions of flare energy, duration, amplitude, and impulse are notably similar between flares detected in the original and denoised CHEOPS light curves, suggesting that the population of recovered flares remains consistent across both datasets. We note a small secondary peak in the flare peak luminosity distribution after denoising, however, that resembles a bimodal pattern that was absent before denoising. We attribute this to the small sample size.

The denoising process led to a higher recovery rate across the entire energy range, with the most significant improvements observed for flares weaker than $10^{29}$ erg. This is consistent with the results of the injection-recovery test, which suggested that the flare recovery rate started to drop at this energy (see Figure \ref{fig:confusion_matrices_evolution2}). As can be seen from the amplitude and duration histograms, the denoising allows us to recover an increased proportion of low amplitude and short duration flares, while those with relative amplitudes higher than $10^{-2}$ and durations longer than $\sim$4 minutes are recovered equally well. This confirms that observing cadence and photometric precision is the main bottleneck for low-energy flare detection. By optimising the S/N of weak flares, the denoising improves one of these factors and facilitates the recovery of a greater proportion of low-amplitude flares. Additionally, denoising appears to increase the flare recovery across most of the impulse range. Contrarily to \cite{Hawley_2014}, we did not find that individual components of complex flares have longer durations than simple flares.

We provide a tentative linear fit to evaluate correlations between components. A power-law relation is visible between the flare duration and recovered energy, with longer flares generally releasing more energy. Similarly, the recovered energy logically scales with amplitude. On the other hand, the recovered energy does not seem to be correlated to the flare impulse, implying that the flare detection and breakdown algorithm efficiently recovers flares of various impulsiveness. A slight trend is visible between the flare amplitude and duration, where the flare duration appears to scale inversely with the amplitude. This might be due to a detection bias and implies that within this energy range, the detection of flares of low amplitude and short durations (i.e. the flares at the lower end of the energy range) is incomplete. This is expected considering the results of the injection-recovery process (see again Figure \ref{fig:confusion_matrices_evolution2}). Finally, as expected, flare peak luminosities scale with flare amplitudes, and FWHMs scale with durations, as these properties act as proxies for each other.

Considering the denoised light curves, flares were detected on 62 out of the 110 stars in our sample. In Figure \ref{fig:flaring_stars_parameters}, we compare the stellar parameters of the flaring stars to the ones of the entire stellar sample. No clear trends are evident in $G_{\mathrm{mag}}$, $T_{\mathrm{eff}}$, radius,  $v$\,sin\,$i$, distance or spectral type. Previous studies have attributed a higher proportion of flaring stars to fast-rotating stars, which are generally younger and possess more active magnetic dynamos than their slower-rotating counterparts \citep{Howard_2020}. This trend has been observed to decline significantly in ultra-fast rotators ($P_{rot} < 0.2$ days), however, which might be due to the high rotation speed that constrains the magnetic field lines in the stellar chromosphere, thereby inhibiting magnetic reconnection \citep{Gunther_2020, Doyle_2022, Ramsay_2022}. Previous large-scale studies have also attributed a higher rate of flaring stars on late Ms compared to their earlier counterparts \citep{Gunther_2020, Pietras_2022}. Our results do not provide sufficient evidence to support any of these hypotheses, likely due to the limited sample size.

Figure \ref{fig:flaring_stars_parameters_per_sptype} shows statistics on the flares separated by spectral subtype. The majority of flares were detected on M1V stars, likely due to the substantial observing time dedicated to AU Mic in our sample. M5V stars produced the fewest flares, likely due to the small number of stars in this group. The mean flare energy appears consistent across spectral subtypes but slightly decreases towards late-type stars. This effect is expected as the average quiescent luminosity decreases with spectral type (a similar trend is visible in Figure \ref{fig:flare_energies}). These mean flare energies remain lower by more than an order of magnitude than the mean flare energy recovered by \cite{Bruno_2024} using combined TESS and CHEOPS data, however, and they are lower by several orders of magnitude than the flare samples observed by TESS \citep{Gunther_2020, Pietras_2022}, Kepler \citep{Davenport_2016, Yang_2017}, and Evryscope \citep{Howard_2019}. When comparing the mean equivalent duration, we find that it tends to increase with later spectral subtypes, suggesting that late M dwarfs produce flares that are more energetic relative to their quiescent luminosities. This trend could be attributed to enhanced magnetic activity in fully convective stars. Flare rates appear to be consistent across spectral subtypes, except for M3V stars, which exhibit a mean of 30 events per day. Most distributions are heavily skewed by a few frequently and intensely flaring stars, however, as indicated by mean values exceeding the third quartile. This also highlights the limited statistical confidence due to the small sample size.

\begin{figure}
\centering
\includegraphics{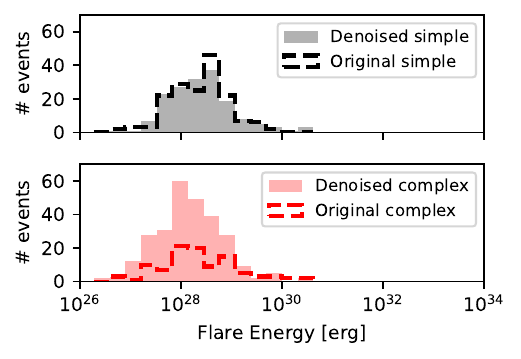}
\caption{Histograms of the number of flares observed per energy bin. The grey shaded histogram represents simple flares in denoised light curves, and the red shaded histogram represents individual components of complex flares. The dashed black and red histograms correspond to the same in the original light curves.}
\label{fig:tess_comparison}
\end{figure}

To better compare the energy range of our flare sample with that of \cite{Bruno_2024}, we display in Figure \ref{fig:tess_comparison} the histogram of the energy distribution of our original and denoised flares, separated into simple flares and individual components of complex flares. First, we find that denoising preserves the number of simple flares while significantly increasing the number of complex flares, suggesting that it helps to break down initially simple flares into several components. Likewise, the number of high-energy complex flares remains similar, indicating that denoising helps to further resolve them into finer components. When comparing the probability density of the flare energies with those in Figure 13 of \cite{Bruno_2024}, the mean flare energy we recovered is lower by more than an order of magnitude for simple and complex flares, which is likely due to the contribution of flares detected by TESS composing the majority of their sample.

\begin{figure}
\centering
\includegraphics{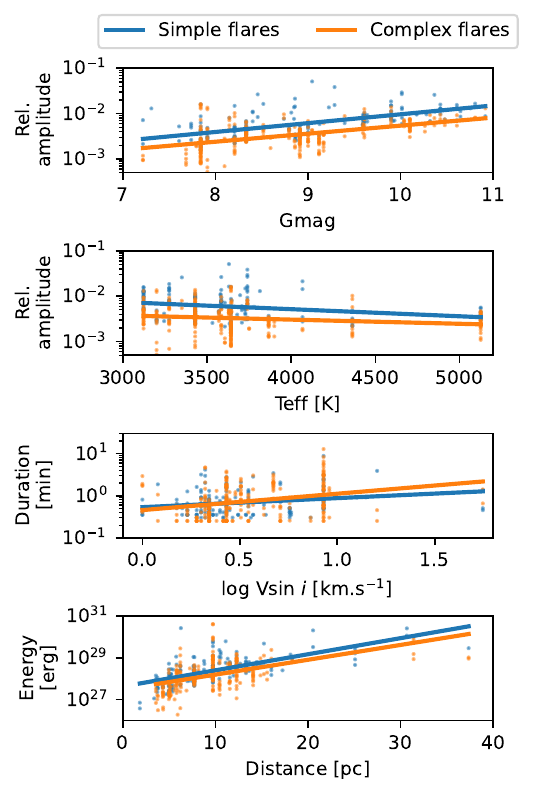}
\caption{Correlations between flare parameters and stellar parameters. The blue and orange data points correspond to simple flares and individual components of complex flares, respectively. Each panel includes a linear fit for reference.}
\label{fig:correlations}
\end{figure}

To explore potential relations between flare parameters and the host star properties, we analysed correlations between flare parameters and those of the corresponding star. We show some of these correlations in Figure \ref{fig:correlations}. A slight exponential trend is observed between the targets' $G$-band magnitude and the relative amplitude of the recovered flares. This trend may be attributed to low-amplitude flares being more easily detectable on brighter stars and suggests that a portion of such flares remain undetected in the noise floor of the light curves of dimmer stars. \citet{Jackman_2021} and \citet{Pietras_2022} found a correlation between the flare amplitude and the effective temperature of the star, with flares detected on cool stars having, on average, higher amplitudes because of the greater contrast between the flaring and quiescent emission. We find a similar trend, although less pronounced, for our simple and complex flares. 
Moreover, a power-law trend is visible between the flare duration and the stellar rotational velocity, indicating that longer flare durations are usually associated with faster-rotating stars. This may be related to fast-rotating stars often having stronger magnetic fields generated by the dynamo effect, which can lead to more intense reconnection events. 
Finally, an exponential trend is observed between the flare energy and the distance of the star. This is likely due to a detection bias, where low-energy flares become increasingly difficult to detect as the distance to the star increases. 
We present all remaining correlations, classified as simple versus complex flares, in Figure~\ref{annex:simpleVScomplex}, and provide the same correlations categorised as flares detected in the original versus denoised light curves in Figure~\ref{annex:originalVSdenoised}. 

\begin{figure*}
\centering
\includegraphics{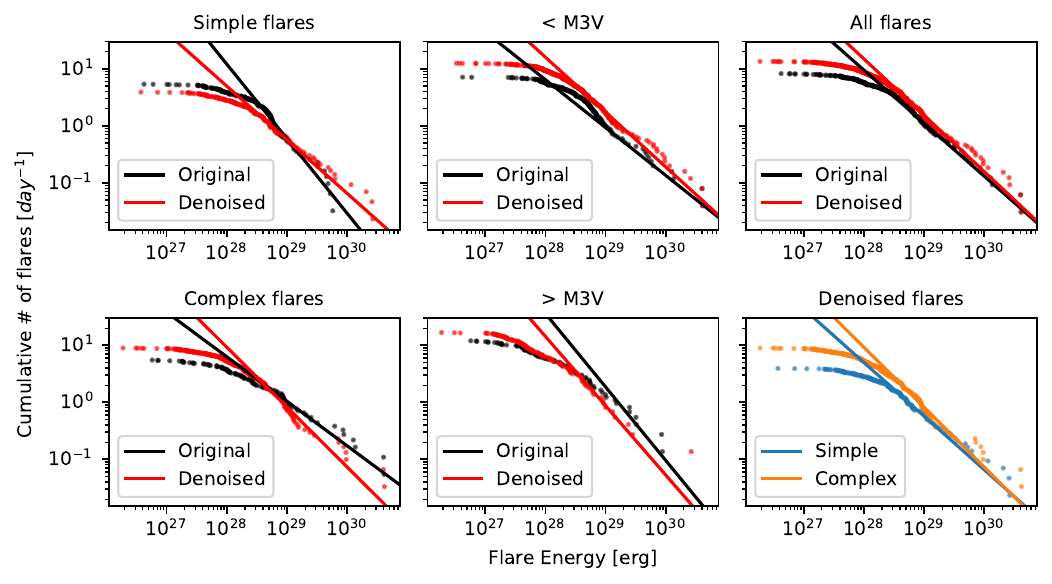}
\caption{Flare frequency distributions for the flares recovered in our original (black) and denoised (red) samples. We separate FFDs into simple flares (top left), individual components of complex flares (bottom left), flares recovered on partially convective stars (top middle), flares recovered on fully convective stars (bottom middle), and the entire flare samples (top right). We also provide a comparison between the simple flares (blue) and individual components of complex flares (orange) recovered in the denoised light curves (bottom right). Solid lines indicate linear fits to the double-logarithmic FFD, extrapolating into regimes not directly observed.}
\label{fig:all_ffds}
\end{figure*}

The relation between the flare energy and occurrence frequency is generally assumed to be characterised by a power-law described as $dN / dE \sim E^{-\alpha}$, where $\alpha$ is the power-law index governing the frequency $dN/dE$ of flaring events with energy $E$. To compare the impact of denoising on our recovered flare samples, we constructed Flare Frequency Distributions (FFDs) for our flare catalogues recovered in original and denoised light curves. The FFDs represent the cumulative rate of flares per day, indicating the frequency of detecting a flare of a given energy or higher. This approach excludes inactive stars from the analysis. In Figure \ref{fig:all_ffds}, we compare the FFDs of flares recovered in original and denoised light curves. For the denoised sample, we also compare FFDs of simple flares and individual components of complex flares.
Following \cite{Yang_2023}, we also separate flares from fully and partially convective stars by setting the separation threshold at M3V. We observe that simple flares are more frequent before denoising, as are individual components of complex flares with energies above $10^{29}$ erg. We attribute this to the denoising facilitating the identification of additional complex flare components, as shown in Figure \ref{fig:morphology} and in the increased percentages in Figure \ref{fig:peaks} and \ref{fig:tess_comparison}. Thus, some simple flares are broken down into multiple components, and high-energy components are split into smaller ones after denoising, explaining the increase in observed frequency. When comparing the full sample of original versus denoised flares, we see that the frequency of recovered flares is higher after denoising across the entire energy range. Even so, the denoising process only marginally extends the lower limit of the detected energy range. 
Similarly, when comparing flares from fully and partially convective stars, the denoising process generally allows us to recover an increased frequency of flares across the entire energy range. This is not the case for flares from fully convective stars above $10^{29}$ erg, which could be caused by the smaller sample size compared to partially convective stars (see the comparison of the number of flares per spectral subtype in the first panel of Figure \ref{fig:flaring_stars_parameters_per_sptype}). Finally, in the denoised light curves, individual components of complex flares appear more frequently at the same energy compared to simple flares. This suggests that highly energetic flares are more likely the result of multiple weaker, sympathetic flares originating from neighbouring active regions, rather than single, powerful events. Consequently, within this energy range, flares appear to occur more frequently in groups rather than as isolated occurrences. 

For each obtained FFD, we determined the corresponding power-law index $\alpha$ by using an MCMC fit based on the Bayesian flare frequency predictor outlined in \cite{Wheatland_2004} and \cite{Ilin_2021}. The fit range began at $4.4 \times 10^{28}$ erg for the original light curves and $1.1 \times 10^{28}$ erg for the denoised light curves, corresponding to the energy from which 90\% of flares are expected to be recovered (see $t_3$ and $t_4$ from Figure \ref{fig:confusion_matrices_evolution2}). 
We performed a Kolmogorov-Smirnov test on each fit to assess whether the power-law assumption should be rejected, using a significance level of 0.05 \citep{Weidner_2009}. In all cases, the power-law assumption was accepted with p-values below 0.05. We compile the obtained $\alpha$ values in Table \ref{tab:alphas}. The obtained values correspond to the higher end of the general $\alpha$ range reported in the literature \citep{Gunther_2020, Pietras_2022, Yang_2023}. Specifically, we find $\alpha = 1.99 \pm 0.10$ for the entire flare population recovered in the denoised light curves. This value is close to the crucial threshold of $\alpha_H>2$, above which smaller flares are thought to play a dominant role in coronal heating \citep{Hudson_1991}.)
Caution is advised when interpreting the obtained $\alpha$ indexes, however, because the portion of the energy range used for the fit is limited and only represents a small subset of events. For the flares detected in the original light curves, only about 34\% had energies above $t_4$ and could be included in the power-law fit. Similarly, for the denoised light curves, just 33\% of flares had energies above $t_3$ and were used in the fit. This highlights that extending the detection limit would allow to improve the detection completeness and obtain a more accurate extension of the FFDs towards lower energies.

\begin{table}
\centering
\caption{Power-law indexes $\alpha$ obtained in Figure~\ref{fig:all_ffds}}
\begin{tabular}{cc}
\hline
Category                                   & $\alpha$        \\ \hline
Simple flares (original)                   & 2.31 $\pm$ 0.12 \\
Simple flares (denoised)                   & 1.94 $\pm$ 0.14 \\
Complex flares (original)                  & 1.78 $\pm$ 0.26 \\
Complex flares (denoised)                  & 2.05 $\pm$ 0.14 \\
< M3V (original)    & 1.84 $\pm$ 0.24 \\
< M3V (denoised)    & 1.96 $\pm$ 0.10 \\
> M3V (original) & 2.29 $\pm$ 0.50 \\
> M3V (denoised) & 2.22 $\pm$ 0.19 \\
All flares (original)                      & 1.94 $\pm$ 0.23 \\
All flares (denoised)                      & 1.99 $\pm$ 0.10 \\ \hline
\end{tabular}
\label{tab:alphas}
\end{table}

A power-law distribution is generally assumed when fitting FFDs in order to determine an $\alpha$ index. A power-law implies that flares are governed by a scale-free process, where flares arise from self-organised criticality. This is the current understanding of magnetic reconnection events, where magnetic loops can accumulate energy and release it in a scale-invariant manner, with a size distribution following a power-law. \cite{Verbeeck_2019} and \cite{Sakurai_2022} suggested, however, that a preliminary step should be to verify that the FFD indeed follows a power-law and could not be better described by an alternative distribution. A log-normal distribution would indicate that high-energy flares result from several smaller independent events combining multiplicatively, rather than from a critical build-up and sudden release of energy. This could happen if flares resulted from many independent magnetic reconnection sites or gradual energy accumulation processes. If flares followed a log-normal distribution, it would suggest that the processes driving flare energy release operate within a characteristic energy range, with fewer instances of extreme events, which could explain the 'truncated power-law' distribution observed by \cite{Howard_2019} at high flare energies.

\begin{figure}
\centering
\includegraphics[width=0.475\textwidth]{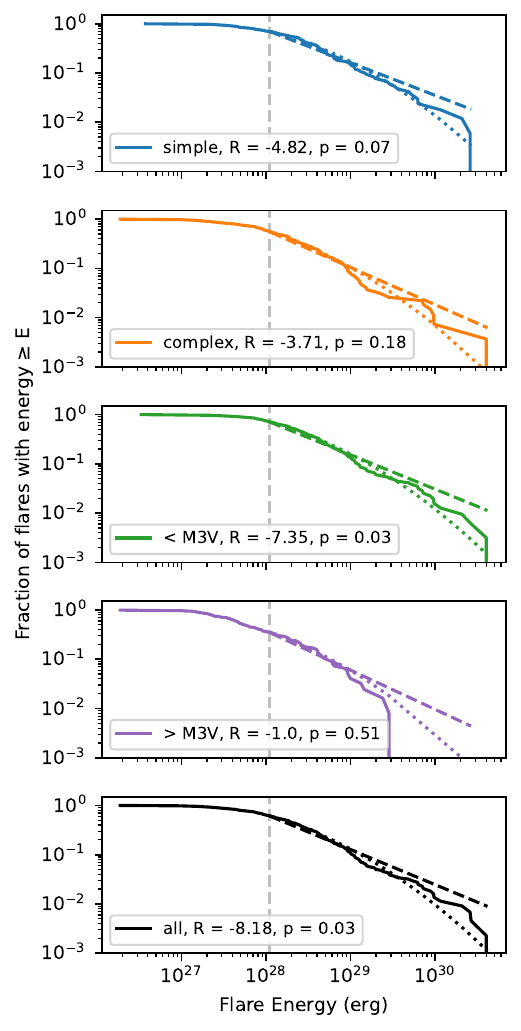}
\caption{Complementary cumulative distribution functions for recovered flare energies in denoised light curves. We separate flares into simple events (blue) and individual components of complex flares (orange), as well as flares from partially (green) and fully convective stars (purple). The full flare sample is displayed in black. Observed distributions, power-law fits, and log-normal fits are displayed with solid lines, dashed lines, and dotted lines, respectively. The grey dashed vertical lines indicate the energy of 90\% recovery ($t_3$) and the beginning of the energy range used for fitting. In each panel, the legend indicates the log-likelihood ratio $R$ between the power-law and log-normal fits and its associated p-value.}
\label{fig:lognormal}
\end{figure}

To test this, we followed the procedure of \cite{Verbeeck_2019} and \cite{Bruno_2024} and compared our obtained FFDs to both a power-law and a log-normal distribution. Figure \ref{fig:lognormal} displays the Complementary Cumulative Distribution Functions (CCDFs) of the flare energies recovered in our denoised light curves, and compares the fitted power-law obtained in Figure \ref{fig:all_ffds} to a log-normal distribution. Again, the fitting range started at the energy of 90\% recovery ($t_3$ = $1.1 \times 10^{28}$ erg). For each denoised category, we identified the best-fitting distribution by calculating the log-likelihood ratio $R$ between the power-law and log-normal fits. A positive $R$ indicates a preference for the power-law, while a negative $R$ favours the log-normal. Our results show that a log-normal distribution better fits the data in all cases, although the p-value of complex flares and flares from fully convective stars is found above 0.05, showing a lack of statistical significance for these subsets. This further emphasises that extending flare observations to lower energies could help clarify the best-fitting distribution for FFDs, which may have important implications for understanding the mechanisms driving stellar flares.

\section{Discussion}
\label{section:discussion}

The use of CHEOPS allowed us to study an unexplored region of the white-light flare energy range in main sequence M dwarf stars. Moreover, we employed a tailored denoising algorithm based on the SWT to optimise the detection rate of low-energy flares. Several potential improvements could extend this work, however. 

The primary limitation encountered in the flare observation stems from the constraints of the detection algorithm employed. The algorithm we used was initially developed for standard flare detection and optimised for TESS and Kepler light curves. Since low-energy flares are generally detected just above the $3\sigma$ threshold, many flares with variations below this level are missed. Lowering the $N_1$ criterion could help capture more low-energy flares, but would significantly increase the false positive rate, thereby affecting the overall flare statistics. We therefore highlight the limitations of a classical sigma-clipping detection method when applied to low-energy flares. Emerging detection algorithms, such as those based on hidden Markov models, use non-stationary time-series and have shown improved efficiency in detecting low-energy flares and accurately measuring their energy \citep{Esquivel_2024, Zimmerman_2024}. Flare detection algorithms employing deep neural networks and random forest classifiers have also demonstrated improved recovery rates and increased sensitivity to low-energy events in Kepler and TESS data, outperforming classical detection techniques \citep{Vida_2021, Lin_2024}. Integrating such algorithms could improve the sensitivity to low-energy flares and increase the confidence in the derived flare statistics.

Another limitation of the flare detection algorithm is the required number of consecutive data points for identifying a flare candidate. To minimise false positives, we set the minimum consecutive data points for detection to $N_3$ = 5. Given that the observing cadence for most light curves we studied is 3 seconds (see Figure \ref{fig:orbit_durations}), this results in a minimum detectable flare duration of 15 seconds. Several studies using observing cadence of 1 second or faster have shown, however, that flares can have durations as brief as a few seconds \citep{Schmitt_2016, Aizawa_2022}. As illustrated in Figure \ref{fig:corner}, flare energy typically scales with duration, which is supported by extensive statistical analyses conducted by TESS and Kepler \citep{Davenport_2016, Howard_2019, Pietras_2022}. Enhancing the observing cadence while conserving the same photometric precision could enable the detection of these shorter-duration, less energetic flares and potentially extend the flare detection toward the lower end of the energy range. CHEOPS can reach an exposure time as fast as 0.001 seconds for very bright stars (V<6) \citep{Broeg_2013, Benz_2021}. Conducting a dedicated observing programme targeting such bright stars could provide high-cadence light curves of active stars, thus decreasing the minimum detectable flare duration. 

An increased cadence would capture more data points, allowing for more precise energy estimates \citep{Clarke_2024}. Denoising low-cadence time-series can cause flare peaks to be treated as outliers and pulled toward the quiescent level, particularly for very short flares with limited data points. This effect leads to a higher proportion of lost energy and increased relative error in energy calculations, as the flare equivalent duration is determined from the area under the curve. Accelerating the observing cadence could help alleviate this issue by providing more data points and minimising the impact of denoising on flare peak values.

Moreover, the flare profile used for the injection and recovery process in Section \ref{subsection:injection_and_recovery} is based on GJ 1243 flare data observed by Kepler \citep{Davenport_2014, Mendoza_2022}. \cite{Seli_2024} identified a correlation between the average flare shape and spectral type, however, which this profile does not account for. Additionally, \cite{Pietras_2022} demonstrated that this model overestimates the decay phase of short-duration flares, while \cite{Dillon_2020} and \cite{Grant_2023} found that low-energy flares tend to exhibit more symmetric profiles with balanced rise and decay times. This suggests that the profile used here may not be ideal for modelling low-energy flares, potentially leading to the selection of suboptimal denoising parameters. As a result, the choice of mother wavelet might be biased toward identifying longer, higher-energy flares while being less effective for shorter, low-energy ones. A more representative flare profile in the injection and recovery process could improve the determination of optimal denoising parameters, requiring a detailed analysis of how flare morphology varies with energy.

Furthermore, the injection and recovery process in Section~\ref{subsection:injection_and_recovery} were limited to simple flares. At high cadence, however, flares often appear as stacked events, adding complexity to their light curve profiles \citep{Davenport_2014, Bruno_2024}. Even low-energy flares can exhibit diverse morphologies, showing the need of separating individual events through improved photometric sensitivity and cadence. A more refined injection-recovery process incorporating both simple and complex flares would ensure that the chosen denoising parameters effectively capture both types. Extending the empirical flare model of \cite{Mendoza_2022} to include well-known morphologies (e.g., 'peak-bump,' 'flat-top') could also help misclassification of complex flare events.

Additionally, recent advancements in wavelet transform applications for time-series denoising offer promising avenues for improvement. Polyharmonic wavelets, which use functions able to represent multiple harmonics, provide better frequency localisation and improved signal decomposition across multiple scales, reducing artefacts in the reconstructed signal \citep{Kouchev_2010}. Although not in an astronomy context, \citep{Ji_2016} and \cite{Alvarado_2023} have also explored hybrid approaches that combine different wavelets or transformation methods to optimise denoising for varying signal complexities. Investigating how these methods impact flare recovery rates compared to standard SWT denoising could provide valuable insights and help refine denoising parameters for optimal performance.

Most importantly, the main limitation of this study is the relatively small sample size, with only 62 flaring stars identified and less than a thousand flaring events catalogued. This constraint arises from the limited data gathered by CHEOPS due to its on-target observing mode. While precise for focused targets, this mode is less effective at quickly monitoring a large number of targets than survey-operating missions like TESS and Kepler. As a result, flare statistics derived from CHEOPS data carry more uncertainty than those from TESS and Kepler, which observed thousands of stars \citep{Davenport_2016, Pietras_2022}. FFDs are generally constructed using hundreds or thousands of flaring events. In our study, however, the average number of flares per star was around 10 for each spectral subtype. This small sample size likely impacts the reliability of the derived FFDs, leading to substantial uncertainties in the calculated $\alpha$ values and possibly explaining the better performance of log-normal fits over power-law fits. This limitation emphasises that, while CHEOPS provides the photometric precision and observing cadence needed for low-energy flare observations, its on-target mode makes it less suited for large-scale flare studies. Future missions combining high photometric precision, fast cadence, and survey-mode capabilities will be crucial for advancing our understanding of low-energy flare statistics.

PLATO (PLAnetary Transits and Oscillations of stars), scheduled for launch at the end of 2026, stands out as the most promising candidate for this matter. Although its primary objective is to observe transits of Earth-like exoplanets around Sun-like stars, it will also collect photometric data on active late-type stars as part of its stellar variability programme. PLATO is anticipated to achieve a photometric precision comparable to CHEOPS, with a fast observing cadence of up to 2 seconds in its high-cadence mode. It will monitor two $49^{\circ} \times 49^{\circ}$ regions of the sky for two continuous years each \citep{Rauer_2024}. The combination of observational capabilities and a survey-mode strategy can significantly increase the detection of low-energy flares and improve the precision of their statistics for late-type stars. Additionally, for the brightest targets, PLATO will provide simultaneous photometry in two bandpasses (red and blue), allowing for direct colour measurements of flares. This will enable more accurate blackbody temperature estimates and extend energy modelling towards low-energy flares, reducing uncertainties in flare energetics. Moreover, the long-term uninterrupted observations from PLATO could help track flare activity cycles and investigate correlations with stellar rotation and magnetic activity over multi-year timescales.

Furthermore, the recently announced TESS Extended Mission 3 presents great capabilities for low-energy flare studies. TESS has been of prime importance in gathering flare statistics in recent years and is set to extend its observing cadence to 2 seconds from 2025 to 2028. While its photometric precision remains lower than those of CHEOPS and PLATO, the enhanced cadence will improve the detection of lower-energy flares compared to previous TESS surveys. This would help bridge the gap between large-scale studies of regular flares and studies of low-energy flares, which remain limited by small sample sizes \citep{Gunther_2020, Pietras_2022, Feinstein_2024}.

Finally, the type of denoising algorithm we presented might improve the recovery of low-energy flares in JWST exoplanet spectra and facilitate the understanding of the impact of flares on exoplanetary atmospheric retrievals \citep{Howard_2023}. During occultations, certain atmospheric elements (e.g. CO$_2$) can produce spectral profiles similar to flare activity, potentially biasing the measurements of emission features of planetary atmospheres \citep{Lustig_2019}. Likewise, flares themselves can mimic the appearance of emission lines from various atmospheric elements, complicating the accurate identification of chemical signatures \citep{Lin_2023}. Flares are known to produce strong chromospheric emission in key spectral lines such as H$_{\alpha}$, CaII H\&K, and HeI 10830 $\AA$, which are also used as tracers for planetary atmospheres \citep{Jensen_2012, Yan_2019, Sanz_2025}. This can lead to false detections of atmospheric species in transmission spectra if the flare signal is not properly accounted for \citep{Konings_2022}. Similarly, in emission spectra, flares can alter the measured stellar flux during a planetary occultation, introducing variability in the observed planetary signal. Additionally, the intense UV and X-ray radiation from flares can heat the upper layers of planetary atmosphere, enhancing atmospheric escape and affecting long-term emission measurements. This effect is particularly important for planets orbiting M dwarfs due to their enhanced activity. Since both transmission and emission spectra can be contaminated by flare-induced spectral features, it is essential to develop techniques to separate true planetary atmospheric signals from stellar activity \citep{Thompson_2024}. Applying a similar denoising algorithm could help differentiate between genuine atmospheric features and transient flare signals, thereby reducing the reduce the risk of misidentifications.

\section{Conclusion}
\label{section:conclusion}
We used CHEOPS to study white-light flares in an unexplored region of the energy range. We employed a denoising algorithm based on the SWT to enhance the detection of low-energy flares in the light curves of a sample of M dwarfs. We optimised the denoising parameters by performing a flare injection and recovery in synthetic light curves that replicated the noise profile of the CHEOPS instrument. The CHEOPS light curves were then denoised using a db18 wavelet and a decomposition level tailored to the number of data points in each visit. Each flare was analysed by a breakdown algorithm to identify substructures. We finally compared the flare populations detected in the original and denoised light curves. 

After denoising, our sample was composed of 291 flaring events, with energies ranging between $3.7 \times 10^{26}$ erg and $8.9 \times 10^{30}$ erg. We identified a portion of $\sim$42\% of flares that had complex structures. 123 complex flares were divided into individual components by using the flare-breakdown algorithm. The denoising process significantly improved the flare detection. It  increased the overall recovery rate by $\sim$35\% and the identification of individual flare components by $\sim$64\%. An improvement in recovery was observed across the entire energy range, but the highest gain was obtained for flares with energies lower than $10^{29}$ erg. We identified several trends between flare parameters that are likely due to a detection bias, which suggests that the flare population we studied is still not entirely recovered at low energies.

The limited sample size prevented us from establishing any clear relation between the flaring frequency and the spectral type, or between the proportion of flaring stars and the spectral type. We observed a trend, however, in which the mean flare energy decreased from early- to late-M stars, while the mean equivalent duration increased. We constructed FFDs and applied power-law fits within the full recovery regime to determine the $\alpha$ indexes. For the complete flare sample, we obtained $\alpha$ = 1.99 $\pm$ 0.10, which is at the higher end of the values typically reported in the literature. We found no statistically significant differences in the $\alpha$ indexes between the distributions of simple and complex flares or between flares from partially and fully convective stars.

We also explored an alternative scenario to describe the observed flare distribution by fitting a log-normal distribution and comparing it to the traditional power-law. The log-likelihood-ratio tests consistently favoured the log-normal fit over the power-law, although statistical confidence was not achieved for all subsets. If flares were indeed better represented by a log-normal distribution, it might indicate that high-energy flares arise from the superposition of multiple weaker sympathetic flares that erupt from nearby active regions. This shift could have strong implications for understanding the flare-formation mechanisms, for the impact of flares on exoplanetary atmospheres, and for assessing long-term stellar activity.

Finally, we reviewed the main limitations of this study. First, the small sample size introduces considerable uncertainty in the derived flare statistics, which leads to a low confidence in the $\alpha$ indexes extracted from the FFDs. Second, the flare-detection algorithm and the model we used to inject flares were both designed for regular flare studies and may bias the denoising parameter selection. This might favour the recovery of regular over low-energy flares. Third, although CHEOPS outperforms most other instruments in photometric precision and observing cadence, which allows it to detect flares with energies corresponding to the upper end of micro-flares, it still cannot capture all low-energy flares. This creates a detection bias in the flare catalogue because the detections are incomplete. We proposed several methods for mitigating these limitations and commented that upcoming missions, such as PLATO and the third mission extension of TESS, could significantly enhance low-energy flare statistics.

\begin{acknowledgements}
The authors thank the anonymous reviewer for their insightful comments, which greatly contributed to improving the manuscript.
This work was partially supported by the Spanish programme MICIN/AEI/10.13039/501100011033 and by the “European Union Next Generation EU/PRTR” through grant PCI2022-135049-2 and by the “ERDF A way of making Europe” by the European Union through grants PID2021-125627OB-C31 and PID2022-136828NB-C41, by the programme Unidad de Excelencia María de Maeztu CEX2019-000918-M to the ICCUB and CEX2020-001058-M to the ICE-CSIC, and by the Generalitat de Catalunya/CERCA programme. I.R. acknowledges financial support from the European Research Council (ERC) under the European Union’s Horizon Europe programme (ERC Advanced Grant SPOTLESS; no. 101140786). 
This research used the new computing cluster Nyx of the ICCUB. 
CHEOPS is an ESA mission in partnership with Switzerland with important contributions to the payload and the ground segment from Austria, Belgium, France, Germany, Hungary, Italy, Portugal, Spain, Sweden, and the United Kingdom. CHEOPS public data analysed in this article are available in the CHEOPS mission archive.
We also used the Data Analysis Center for Exoplanets (DACE), dedicated to extrasolar planets data visualisation, exchange, and analysis. DACE is a platform of the Swiss National Centre of Competence in Research (NCCR) PlanetS, federating the Swiss expertise in Exoplanet research. 
Additional resources include the SIMBAD database and VizieR catalog access tool from CDS, Strasbourg Astronomical Observatory, France \citep{Wenger_2000, Ochsenbein_2000}. \\
Software: \texttt{Python} \citep{VanRossum_2009}, \texttt{AltaiPony} \citep{Ilin_2022}, \texttt{AstroPy} \citep{Astropy_2022}, \texttt{AstroQuery} \citep{Ginsburg_2019}, \texttt{Matplotlib} \citep{Hunter_2007}, \texttt{NumPy} \citep{Harris_2020}, \texttt{Pandas} \citep{McKinney_2010}, \texttt{PIPE} \citep{Brandeker_2024}, \texttt{Powerlaw} \citep{Alstott_2014}, \texttt{PyCHEOPS} \citep{Maxted_2023}, \texttt{PyWavelets} \citep{Lee_2019}, and \texttt{SciPy} \citep{Virtanen_2020}.

\end{acknowledgements}

\bibliographystyle{aa}
\bibliography{aanda}

\onecolumn
\begin{appendix}
\section*{Appendix A: Target list.}
\renewcommand{\thetable}{A.\arabic{table}}
\begin{longtable}{lccccccc}
\caption{\label{targetlist} List of targets and associated parameters.}\\
\hline
Name & Spectral type & Gmag & $\mathrm{T}_{\mathrm{eff}}$ [K] & $\mathrm{Vsini}$ [km.s$^{-1}$] & Distance [pc] & $\mathrm{Radius}$ [$\mathrm{R}_\odot$] & Obs. time [h] \\
\hline
\endfirsthead
\caption{continued.}\\
\hline
Name & Spectral type & Gmag & $\mathrm{T}_{\mathrm{eff}}$ [K] & $\mathrm{Vsini}$ [km.s$^{-1}$] & Distance [pc] & $\mathrm{Radius}$ [$\mathrm{R}_\odot$] & Obs. time [h] \\
\hline
\endhead
\hline
\endfoot
2MASS J03413724+5513068 & M2V           & 10.55 & 4050.00      & 4.5                  & 35.85 & 0.663  & 0.83         \\
2MASS J06144242+4727346 & M0V           & 10.81 & 3739.48      & 7.3                  & 37.35 & 0.619  & 3.70         \\
2MASS J06192947+1357031 & M0V           & 10.01 & 3739.48      & 1.0                  & 25.08 & 0.610  & 3.51         \\
2MASS J09304457+0019214 & M3V           & 10.49 & 3275.05      & 1.6                  & 9.90  & 0.323  & 3.18         \\
2MASS J11421839+2301365 & M0V           & 10.83 & 3739.48      & 1.0                  & 30.71 & 0.546  & 1.62         \\
2MASS J11474440+0048164 & M3V           & 9.59  & 3122.25      & 3.7                  & 3.37  & 0.210  & 4.29         \\
2MASS J13314666+2916368 & M4V           & 10.61 & 3122.25      & 55.8                 & 18.29 & 0.539  & 7.10        \\
2MASS J20103444+0632140 & M4V           & 10.92 & 3122.25      & 1.0                  & 16.03 & 0.421  & 6.92         \\
2MASS J21462206+3813047 & M5V           & 10.82 & 2971.27      & 1.4                  & 7.04  & 0.214  & 1.61         \\
2MASS J22232904+3227334 & M0V           & 10.37 & 3350.00      & 8.5                  & 15.23 & 0.593  & 12.65        \\
2MASS J23415498+4410407 & M5V           & 10.37 & 2971.27      & 2.5                  & 3.16  & 0.178  & 2.43         \\
AD Leo                  & M3V           & 8.21  & 4363.00      & 3.5                  & 4.97  & 0.422  & 13.18        \\
AU Mic                  & M1V           & 7.84  & 3642.00      & 8.5                  & 9.72  & 0.698  & 215.80       \\
BD+33 1505              & M0V           & 9.35  & 3619.00      & 3.7                  & 18.22 & 0.598  & 5.30        \\
BD-02 2198              & M1V           & 9.12  & 3866.00      & 3.2                  & 14.07 & 0.577  & 5.72        \\
BX Cet                  & M2V           & 10.32 & 3275.05      & 3.0                  & 7.22  & 0.279  & 0.94         \\
CE Boo                  & M0V           & 9.13  & 3780.00      & 4.3                  & 9.93  & 0.477  & 3.14         \\
EE Leo                  & M4V           & 10.28 & 3122.25      & 2.6                  & 6.97  & 0.293  & 5.65         \\
EG Cam                  & M0V           & 9.41  & 3739.48      & 2.3                  & 13.49 & 0.513  & 1.48         \\
EQ Peg                  & M4V           & 9.04  & 3630.00      & 16.0                 & 6.26  & 0.513  & 3.43         \\
EV Lac                  & M4V           & 9.00  & 3122.25      & 5.1                  & 5.05  & 0.337  & 7.95        \\
G 168-31                & M3V           & 10.98 & 3429.20      & 1.1                  & 36.91 & 0.655  & 6.49         \\
G 214-14                & M0V           & 10.38 & 3739.48      & 1.7                  & 23.71 & 0.513  & 5.18         \\
G 234-57                & M1V           & 10.46 & 3429.20      & 2.0                  & 21.05 & 0.400  & 3.26         \\
G 32-5                  & M4V           & 11.40 & 3122.25      & 5.5                  & 12.21 & 0.269  & 3.12         \\
G 99-49                 & M3V           & 9.90  & 3275.05      & 5.7                  & 5.21  & 0.261  & 21.53        \\
GJ 1                    & M2V           & 7.68  & 3429.20      & 2.8                  & 4.35  & 0.396  & 8.54        \\
GJ 1074                 & M0V           & 10.15 & 3584.18      & 4.0                  & 21.11 & 0.537  & 6.64        \\
GJ 1105                 & M4V           & 10.67 & 3275.05      & 1.9                  & 8.84  & 0.294  & 2.63         \\
GJ 15 A                 & M2V           & 7.22  & 3605.50      & 3.7                  & 3.56  & 0.406  & 15.87        \\
GJ 176                  & M2V           & 9.00  & 3679.00      & 12.6                 & 9.47  & 0.487  & 7.76        \\
GJ 180                  & M2V           & 9.93  & 3275.05      & 1.7                  & 11.94 & 0.413  & 14.15        \\
GJ 184                  & M0V           & 9.21  & 3739.48      & 3.5                  & 13.86 & 0.530  & 4.15         \\
GJ 2                    & M2V           & 9.08  & 3875.00      & 1.8                  & 11.50 & 0.515  & 3.29         \\
GJ 205                  & M1V           & 7.10  & 3731.20      & 3.3                  & 5.70  & 0.561  & 22.31        \\
GJ 2066                 & M0V           & 9.12  & 3429.20      & 1.9                  & 8.94  & 0.443  & 3.73         \\
GJ 229                  & M1V           & 7.31  & 3814.00      & 3.1                  & 5.76  & 0.549  & 56.05        \\
GJ 26                   & M1V           & 10.05 & 3429.20      & 2.2                  & 12.67 & 0.430  & 2.49         \\
GJ 273                  & M4V           & 8.59  & 3275.05      & 2.2                  & 3.79  & 0.316  & 23.02        \\
GJ 317                  & M4V           & 10.75 & 3275.05      & 2.8                  & 15.20 & 0.427  & 6.63        \\
GJ 328                  & M0V           & 9.29  & 3739.48      & 3.4                  & 20.54 & 0.651  & 8.12        \\
GJ 3323                 & M4V           & 10.65 & 3122.25      & 2.3                  & 5.38  & 0.186  & 13.11        \\
GJ 358                  & M0V           & 9.63  & 3275.05      & 1.6                  & 9.60  & 0.423  & 3.60         \\
GJ 3649                 & M1V           & 9.88  & 3584.18      & 1.9                  & 16.68 & 0.529  & 1.03         \\
GJ 382                  & M0V           & 8.33  & 3429.20      & 2.2                  & 7.70  & 0.510  & 35.50        \\
GJ 3822                 & M1V           & 9.83  & 3584.18      & 3.5                  & 20.34 & 0.581  & 2.90         \\
GJ 399                  & M1V           & 10.26 & 3429.20      & 1.7                  & 15.58 & 0.466  & 9.97        \\
GJ 3997                 & M1V           & 9.64  & 3739.48      & 2.7                  & 13.63 & 0.480  & 15.21        \\
GJ 408                  & M2V           & 8.97  & 3122.25      & 2.1                  & 6.75  & 0.390  & 5.90        \\
GJ 4092                 & M0V           & 10.12 & 3739.48      & 2.7                  & 28.23 & 0.630  & 16.53        \\
GJ 422                  & M4V           & 10.48 & 3275.05      & 1.2                  & 12.67 & 0.370  & 4.04        \\
GJ 433                  & M0V           & 8.89  & 3616.00      & 1.3                  & 9.07  & 0.469  & 15.03        \\
GJ 436                  & M1V           & 9.57  & 3416.00      & 1.7                  & 9.76  & 0.425  & 1.62         \\
GJ 450                  & M1V           & 8.85  & 3584.18      & 5.8                  & 8.76  & 0.460  & 20.58        \\
GJ 47                   & M2V           & 9.84  & 4104.00      & 2.0                  & 10.52 & 0.390  & 1.55         \\
GJ 49                   & M2V           & 8.66  & 4055.50      & 2.9                  & 9.86  & 0.540  & 9.60        \\
GJ 494                  & M0V           & 8.91  & 3899.50      & 9.1                  & 11.51 & 0.563  & 16.71        \\
GJ 514                  & M1V           & 8.21  & 3727.00      & 1.9                  & 7.62  & 0.503  & 22.00        \\
GJ 521                  & M2V           & 9.40  & 3584.18      & 2.9                  & 13.37 & 0.498  & 0.98         \\
GJ 526                  & M2V           & 7.61  & 3634.00      & 2.4                  & 5.44  & 0.482  & 16.27        \\
GJ 536                  & M0V           & 8.86  & 4067.00      & 1.7                  & 10.41 & 0.508  & 9.49        \\
GJ 552                  & M1V           & 9.72  & 3429.20      & 2.6                  & 14.25 & 0.503  & 3.46         \\
GJ 581                  & M1V           & 9.41  & 3442.00      & 1.8                  & 6.30  & 0.330  & 10.13        \\
GJ 588                  & M3V           & 8.27  & 3429.20      & 1.8                  & 5.92  & 0.460  & 11.72        \\
GJ 606                  & M0V           & 9.59  & 3584.18      & 2.0                  & 13.29 & 0.487  & 11.34        \\
GJ 628                  & M3V           & 8.79  & 3570.00      & 1.5                  & 4.31  & 0.322  & 20.84        \\
GJ 649                  & M2V           & 8.82  & 3696.33      & 2.1                  & 10.38 & 0.517  & 12.58        \\
GJ 65                   & M6V           & 10.51 & 2971.27      & 26.4                 & 2.72  & 0.165  & 2.00        \\
GJ 674                  & M3V           & 8.33  & 3275.05      & 1.8                  & 4.55  & 0.365  & 13.71        \\
GJ 676 A                & M0V           & 8.87  & 3739.48      & 2.6                  & 16.03 & 0.649  & 7.19        \\
GJ 686                  & M1V           & 8.74  & 3584.18      & 2.9                  & 8.16  & 0.442  & 24.90        \\
GJ 699                  & M1V           & 8.20  & 3244.67      & 2.5                  & 1.83  & 0.194  & 52.41        \\
GJ 70                   & M1V           & 9.90  & 3429.20      & 2.0                  & 11.32 & 0.408  & 5.34        \\
GJ 701                  & M0V           & 8.52  & 3630.00      & 1.9                  & 7.73  & 0.465  & 42.92        \\
GJ 731                  & M0V           & 9.38  & 3739.48      & 2.7                  & 15.21 & 0.539  & 17.27        \\
GJ 740                  & M1V           & 8.46  & 3584.18      & 2.3                  & 11.11 & 0.588  & 31.53        \\
GJ 752 A                & M3V           & 8.10  & 3275.05      & 2.7                  & 5.91  & 0.473  & 36.39        \\
GJ 83.1                 & M5V           & 10.67 & 3122.25      & 2.6                  & 4.47  & 0.180  & 8.20        \\
GJ 832                  & M2V           & 7.74  & 3707.00      & 2.0                  & 4.97  & 0.442  & 13.10        \\
GJ 846                  & M0V           & 8.40  & 3580.00      & 3.1                  & 10.55 & 0.574  & 42.89        \\
GJ 849                  & M0V           & 9.22  & 3275.05      & 1.7                  & 8.80  & 0.464  & 33.19        \\
GJ 876                  & M3V           & 8.88  & 3532.00      & 2.5                  & 4.68  & 0.352  & 19.91        \\
GJ 880                  & M1V           & 7.79  & 3750.00      & 2.4                  & 6.87  & 0.550  & 14.08        \\
GJ 908                  & M1V           & 8.15  & 3646.00      & 2.6                  & 5.90  & 0.417  & 17.19        \\
GJ 9122 B               & M0V           & 9.92  & 3739.48      & 3.6                  & 17.24 & 0.523  & 1.52         \\
GJ 9404                 & M0V           & 9.87  & 3739.48      & 2.6                  & 23.90 & 0.626  & 0.80         \\
GJ 9793                 & M0V           & 10.04 & 3739.48      & 1.0                  & 31.40 & 0.692  & 10.89        \\
Gl 799B                 & M4V           & 9.59  & 3123.00      & 10.2                 & 9.83  & 0.692  & 2.78         \\
Gl 841 A                & M2V           & 9.40  & 3429.20      & 4.2                  & 14.86 & 0.608  & 10.82        \\
HD 154363B              & M1V           & 9.17  & 3584.18      & 2.7                  & 10.46 & 0.463  & 12.01        \\
HD 233153               & M1V           & 8.91  & 5125.96      & 2.7                  & 12.28 & 0.555  & 26.30        \\
HD 265866               & M1V           & 8.86  & 3275.05      & 1.7                  & 5.58  & 0.368  & 15.94        \\
HD 50281B               & M0V           & 9.09  & 4763.86      & 3.9                  & 8.74  & 0.442  & 6.62        \\
HD 79211                & M0V           & 7.05  & 3870.00      & 2.9                  & 6.33  & 0.586  & 43.96        \\
HD 95735                & M2V           & 6.55  & 3563.50      & 7.3                  & 2.55  & 0.389  & 12.65        \\
HIP 57050               & M4V           & 10.58 & 3122.25      & 1.8                  & 11.02 & 0.359  & 2.54         \\
HIP 79431               & M1V           & 10.24 & 3275.05      & 1.0                  & 14.54 & 0.479  & 6.55        \\
LHS 3432                & M0V           & 9.80  & 3429.20      & 4.3                  & 8.82  & 0.336  & 8.11        \\
LP 609-71               & M1V           & 9.61  & 3429.20      & 2.7                  & 11.54 & 0.485  & 6.98        \\
LP 672-42               & M3V           & 10.81 & 3275.05      & 1.5                  & 13.44 & 0.372  & 1.39         \\
MCC 549                 & M0V           & 10.28 & 3739.48      & 19.1                 & 38.80 & 0.815  & 7.72        \\
Proxima Centauri        & M4V           & 8.95  & 2990.50      & 2.6                  & 1.30  & 0.154  & 5.86        \\
Ross 733                & M4V           & 10.37 & 3122.25      & 14.0                 & 18.10 & 0.519  & 2.97         \\
TYC 1313-1482-1         & M0V           & 10.27 & 3739.48      & 1.0                  & 46.23 & 0.870  & 3.34         \\
TYC 4902-210-1          & M0V           & 10.01 & 3739.48      & 1.6                  & 30.67 & 0.706  & 6.58        \\
V 1054 Oph              & M3V           & 7.91  & 3200.00      & 2.1                  & 6.20  & 0.533  & 15.33        \\
V1352 Ori               & M3V           & 10.10 & 3122.25      & 4.7                  & 5.79  & 0.249  & 2.66         \\
VV Lyn                  & M2V           & 10.47 & 3429.20      & 4.6                  & 11.87 & 0.518  & 18.25        \\
Wolf 906                & M1V           & 10.17 & 3429.20      & 1.7                  & 14.46 & 0.461  & 5.61         \\
YZ Ceti                 & M5V           & 10.43 & 3122.25      & 2.2                  & 3.71  & 0.168  & 29.67    
\label{target_list}
\end{longtable}

\newpage

\section*{Appendix B: Wavelet list.}
\setcounter{table}{0}
\renewcommand{\thetable}{B.\arabic{table}}
\begin{longtable}{llccccc}
\caption{\label{wavelet_info} Properties of the wavelets used in Section \ref{subsection:denoising}.}\\
\hline
Family name          & Wavelet name & Vanishing moments & Filters length & Orthogonal & Biorthogonal & Symmetry       \\ \hline
\endfirsthead
\caption{continued.}\\
\hline
Family name          & Wavelet name & Vanishing moments & Filters length & Orthogonal & Biorthogonal & Symmetry       \\ \hline
\endhead
\endfoot
Biorthogonal         & bior1.1      & 1                 & 2              & False      & True         & symmetric      \\
                     & bior1.3      & 1                 & 6              & False      & True         & symmetric      \\
                     & bior1.5      & 1                 & 10             & False      & True         & symmetric      \\
                     & bior2.2      & 2                 & 6              & False      & True         & symmetric      \\
                     & bior2.4      & 2                 & 10             & False      & True         & symmetric      \\
                     & bior2.6      & 2                 & 14             & False      & True         & symmetric      \\
                     & bior2.8      & 2                 & 18             & False      & True         & symmetric      \\
                     & bior3.1      & 3                 & 4              & False      & True         & symmetric      \\
                     & bior3.3      & 3                 & 8              & False      & True         & symmetric      \\
                     & bior3.5      & 3                 & 12             & False      & True         & symmetric      \\
                     & bior3.7      & 3                 & 16             & False      & True         & symmetric      \\
                     & bior3.9      & 3                 & 20             & False      & True         & symmetric      \\
                     & bior4.4      & 4                 & 10             & False      & True         & symmetric      \\
                     & bior5.5      & 5                 & 12             & False      & True         & symmetric      \\
                     & bior6.8      & 6                 & 18             & False      & True         & symmetric      \\
Coiflets             & coif1        & 2                 & 6              & True       & True         & near symmetric \\
                     & coif2        & 4                 & 12             & True       & True         & near symmetric \\
                     & coif3        & 6                 & 18             & True       & True         & near symmetric \\
                     & coif4        & 8                 & 24             & True       & True         & near symmetric \\
                     & coif5        & 10                & 30             & True       & True         & near symmetric \\
                     & coif6        & 12                & 36             & True       & True         & near symmetric \\
                     & coif7        & 14                & 42             & True       & True         & near symmetric \\
                     & coif8        & 16                & 48             & True       & True         & near symmetric \\
                     & coif9        & 18                & 54             & True       & True         & near symmetric \\
                     & coif10       & 20                & 60             & True       & True         & near symmetric \\
                     & coif11       & 22                & 66             & True       & True         & near symmetric \\
                     & coif12       & 24                & 72             & True       & True         & near symmetric \\
                     & coif13       & 26                & 78             & True       & True         & near symmetric \\
                     & coif14       & 28                & 84             & True       & True         & near symmetric \\
                     & coif15       & 30                & 90             & True       & True         & near symmetric \\
                     & coif16       & 32                & 96             & True       & True         & near symmetric \\
                     & coif17       & 34                & 102            & True       & True         & near symmetric \\
Daubechies           & db1          & 1                 & 2              & True       & True         & asymmetric     \\
                     & db2          & 2                 & 4              & True       & True         & asymmetric     \\
                     & db3          & 3                 & 6              & True       & True         & asymmetric     \\
                     & db4          & 4                 & 8              & True       & True         & asymmetric     \\
                     & db5          & 5                 & 10             & True       & True         & asymmetric     \\
                     & db6          & 6                 & 12             & True       & True         & asymmetric     \\
                     & db7          & 7                 & 14             & True       & True         & asymmetric     \\
                     & db8          & 8                 & 16             & True       & True         & asymmetric     \\
                     & db9          & 9                 & 18             & True       & True         & asymmetric     \\
                     & db10         & 10                & 20             & True       & True         & asymmetric     \\
                     & db11         & 11                & 22             & True       & True         & asymmetric     \\
                     & db12         & 12                & 24             & True       & True         & asymmetric     \\
                     & db13         & 13                & 26             & True       & True         & asymmetric     \\
                     & db14         & 14                & 28             & True       & True         & asymmetric     \\
                     & db15         & 15                & 30             & True       & True         & asymmetric     \\
                     & db16         & 16                & 32             & True       & True         & asymmetric     \\
                     & db17         & 17                & 34             & True       & True         & asymmetric     \\
                     & db18         & 18                & 36             & True       & True         & asymmetric     \\
                     & db19         & 19                & 38             & True       & True         & asymmetric     \\
                     & db20         & 20                & 40             & True       & True         & asymmetric     \\
                     & db21         & 21                & 42             & True       & True         & asymmetric     \\
                     & db22         & 22                & 44             & True       & True         & asymmetric     \\
                     & db23         & 23                & 46             & True       & True         & asymmetric     \\
                     & db24         & 24                & 48             & True       & True         & asymmetric     \\
                     & db25         & 25                & 50             & True       & True         & asymmetric     \\
                     & db26         & 26                & 52             & True       & True         & asymmetric     \\
                     & db27         & 27                & 54             & True       & True         & asymmetric     \\
                     & db28         & 28                & 56             & True       & True         & asymmetric     \\
                     & db29         & 29                & 58             & True       & True         & asymmetric     \\
                     & db30         & 30                & 60             & True       & True         & asymmetric     \\
                     & db31         & 31                & 62             & True       & True         & asymmetric     \\
                     & db32         & 32                & 64             & True       & True         & asymmetric     \\
                     & db33         & 33                & 66             & True       & True         & asymmetric     \\
                     & db34         & 34                & 68             & True       & True         & asymmetric     \\
                     & db35         & 35                & 70             & True       & True         & asymmetric     \\
                     & db36         & 36                & 72             & True       & True         & asymmetric     \\
                     & db37         & 37                & 74             & True       & True         & asymmetric     \\
                     & db38         & 38                & 76             & True       & True         & asymmetric     \\
Discrete Meyer       & dmey         &         1          & 62             & True       & True         & symmetric      \\
Haar                 & haar         & 1                 & 2              & True       & True         & asymmetric     \\
Reverse biorthogonal & rbio1.1      & 1                 & 2              & False      & True         & symmetric      \\
                     & rbio1.3      & 1                 & 6              & False      & True         & symmetric      \\
                     & rbio1.5      & 1                 & 10             & False      & True         & symmetric      \\
                     & rbio2.2      & 2                 & 6              & False      & True         & symmetric      \\
                     & rbio2.4      & 2                 & 10             & False      & True         & symmetric      \\
                     & rbio2.6      & 2                 & 14             & False      & True         & symmetric      \\
                     & rbio2.8      & 2                 & 18             & False      & True         & symmetric      \\
                     & rbio3.1      & 3                 & 4              & False      & True         & symmetric      \\
                     & rbio3.3      & 3                 & 8              & False      & True         & symmetric      \\
                     & rbio3.5      & 3                 & 12             & False      & True         & symmetric      \\
                     & rbio3.7      & 3                 & 16             & False      & True         & symmetric      \\
                     & rbio3.9      & 3                 & 20             & False      & True         & symmetric      \\
                     & rbio4.4      & 4                 & 10             & False      & True         & symmetric      \\
                     & rbio5.5      & 5                 & 12             & False      & True         & symmetric      \\
                     & rbio6.8      & 6                 & 18             & False      & True         & symmetric      \\
Symlets              & sym2         & 2                 & 4              & True       & True         & near symmetric \\
                     & sym3         & 3                 & 6              & True       & True         & near symmetric \\
                     & sym4         & 4                 & 8              & True       & True         & near symmetric \\
                     & sym5         & 5                 & 10             & True       & True         & near symmetric \\
                     & sym6         & 6                 & 12             & True       & True         & near symmetric \\
                     & sym7         & 7                 & 14             & True       & True         & near symmetric \\
                     & sym8         & 8                 & 16             & True       & True         & near symmetric \\
                     & sym9         & 9                 & 18             & True       & True         & near symmetric \\
                     & sym10        & 10                & 20             & True       & True         & near symmetric \\
                     & sym11        & 11                & 22             & True       & True         & near symmetric \\
                     & sym12        & 12                & 24             & True       & True         & near symmetric \\
                     & sym13        & 13                & 26             & True       & True         & near symmetric \\
                     & sym14        & 14                & 28             & True       & True         & near symmetric \\
                     & sym15        & 15                & 30             & True       & True         & near symmetric \\
                     & sym16        & 16                & 32             & True       & True         & near symmetric \\
                     & sym17        & 17                & 34             & True       & True         & near symmetric \\
                     & sym18        & 18                & 36             & True       & True         & near symmetric \\
                     & sym19        & 19                & 38             & True       & True         & near symmetric \\
                     & sym20        & 20                & 40             & True       & True         & near symmetric \\ \hline
\label{wavelet_list}
\end{longtable}

\newpage
\section*{Appendix C: Flare recovery heatmaps.}
\renewcommand{\thefigure}{C.\arabic{figure}}

\begin{figure*}[h]
\centering
\includegraphics[width=0.85\textwidth]{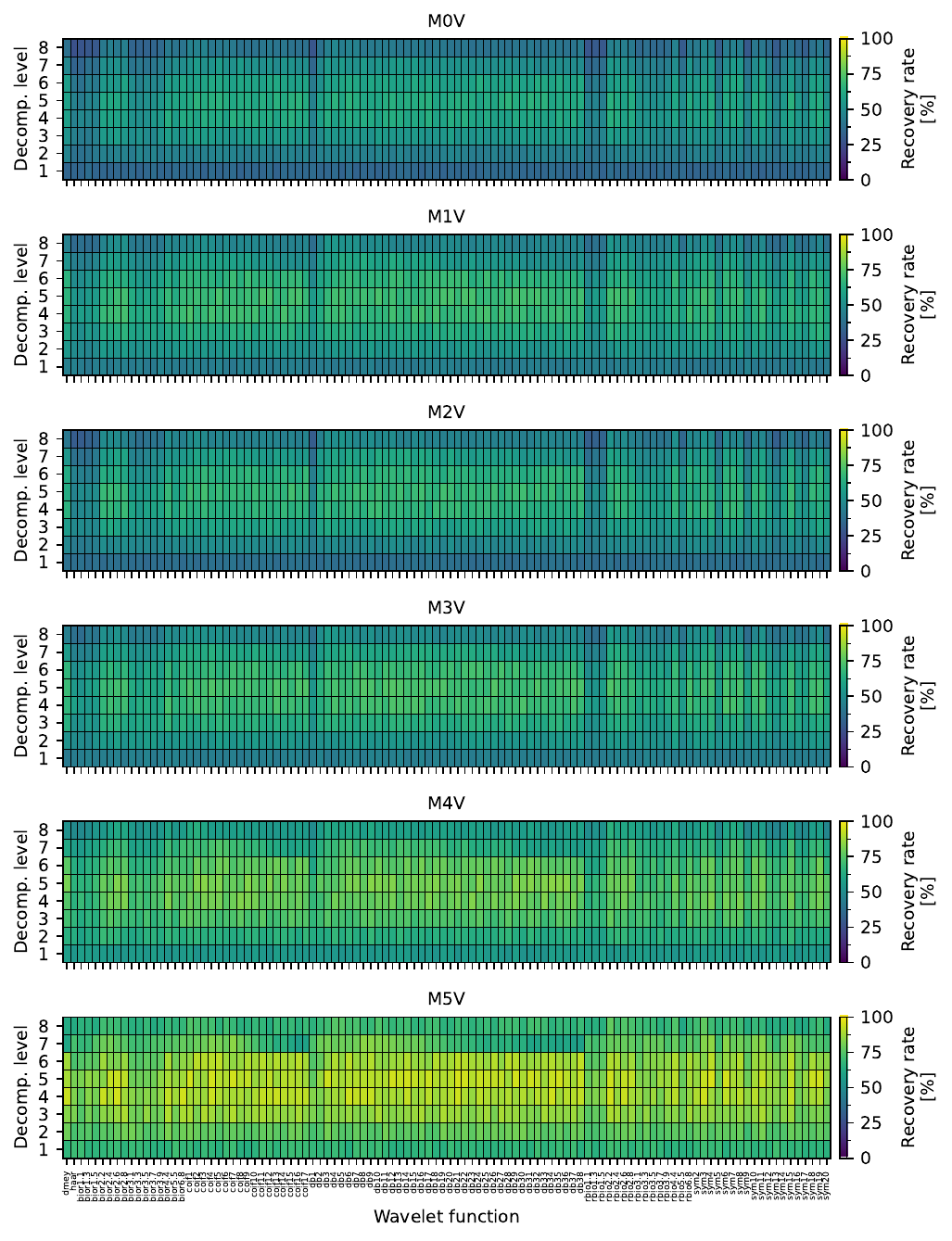}
\caption{Flare recovery rate as a function of the selected mother wavelet and decomposition
level, separated per spectral subtype.}
\label{annex:recovery_rates}
\end{figure*}

\newpage
\section*{Appendix D: Additional flare detections.}
\setcounter{figure}{0}
\renewcommand{\thefigure}{D.\arabic{figure}}

\begin{figure*}[h!]
\centering
\begin{subfigure}{\textwidth}
\centering
\includegraphics{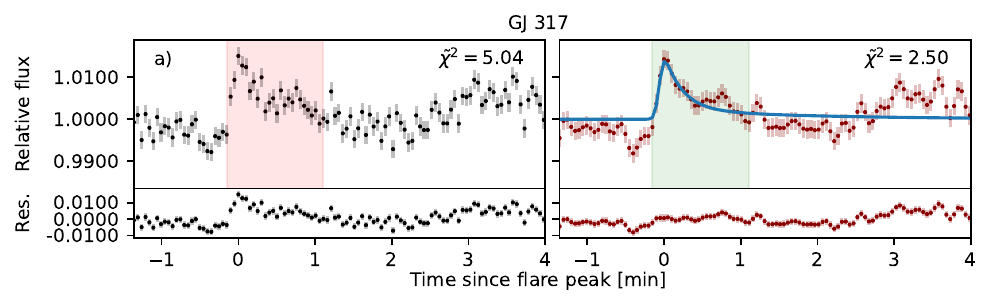}
\end{subfigure}

\begin{subfigure}{\textwidth}
\centering
\includegraphics{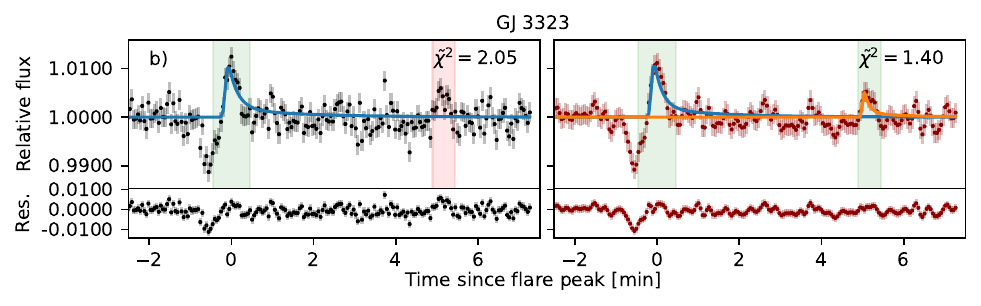}
\end{subfigure}

\begin{subfigure}{\textwidth}
\centering
\includegraphics{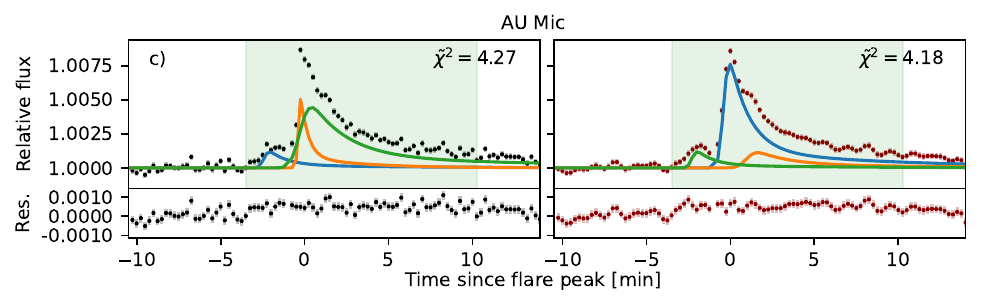}
\end{subfigure}

\begin{subfigure}{\textwidth}
\centering
\includegraphics{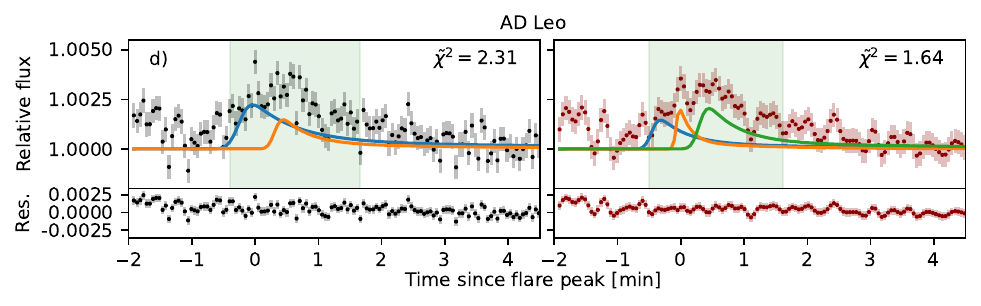}
\end{subfigure}

\caption{Additional flare detections similar to Figure \ref{fig:grid}. This figure shows two pre-dip flares (panels a and b) and two QPP candidates (panels c and d).}
\label{annex:dip+qpp}
\end{figure*}

\newpage

\begin{figure*}[h!]
\centering
\begin{subfigure}{\textwidth}
\centering
\includegraphics{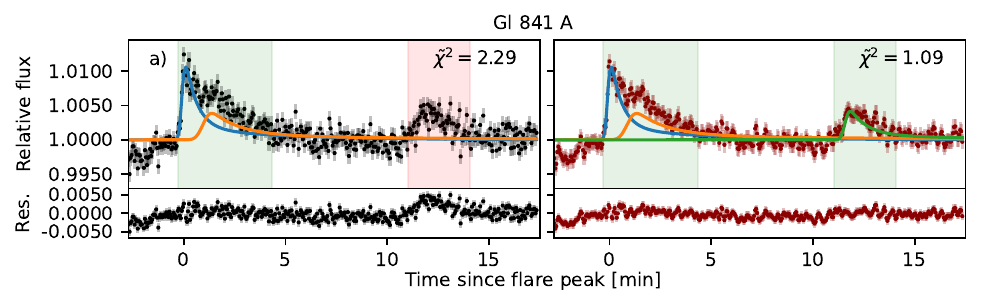}
\end{subfigure}

\begin{subfigure}{\textwidth}
\centering
\includegraphics{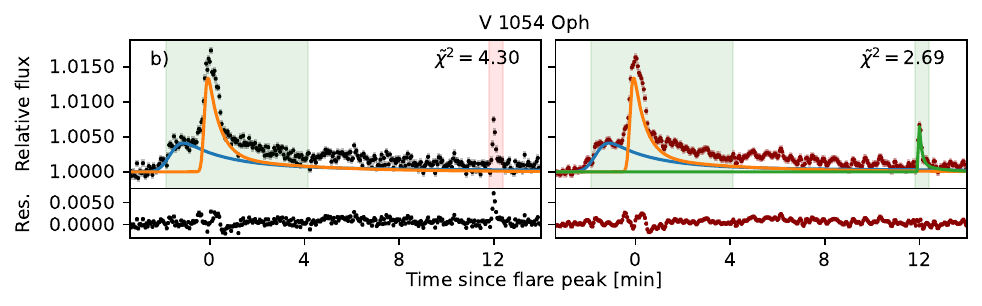}
\end{subfigure}

\begin{subfigure}{\textwidth}
\centering
\includegraphics{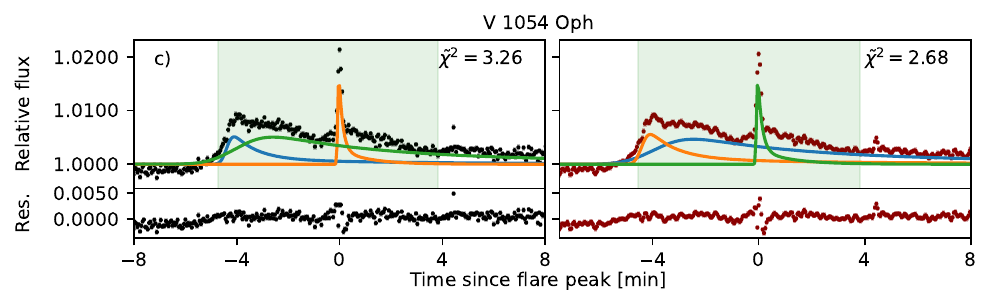}
\end{subfigure}

\begin{subfigure}{\textwidth}
\centering
\includegraphics{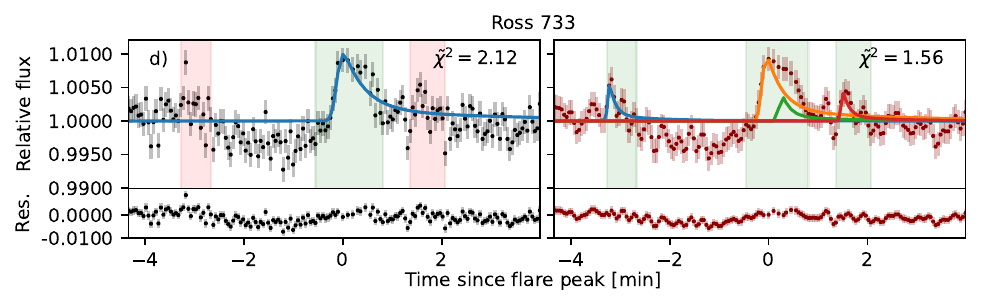}
\end{subfigure}

\caption{Additional flare detections similar to Figure \ref{fig:grid}. Panel a) shows a flare with a 'peak-bump' profile, panels b) and c) show flares with a 'bump-peak' profile, and panel d) shows a flare with a 'flat-top' profile.}
\label{annex:morphologies}
\end{figure*}

\newpage

\section*{Appendix E: Flare-star parameters correlations.}
\setcounter{figure}{0}
\renewcommand{\thefigure}{E.\arabic{figure}}

\begin{figure*}[h]
\centering
\includegraphics{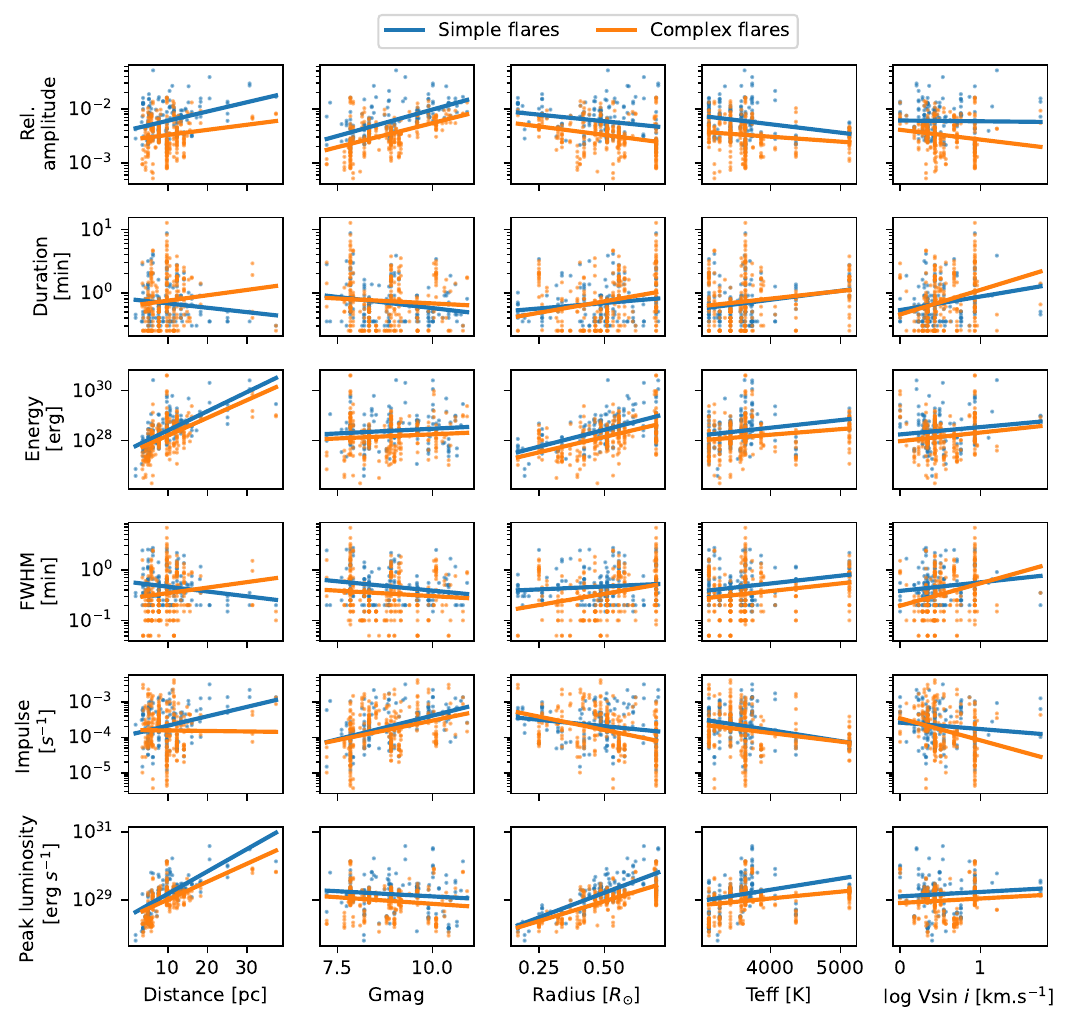}
\caption{Correlations between the denoised flare parameters and the ones of its star, for simple flares (blue) and individual components of complex flares (orange). Linear fits are shown for indication.}
\label{annex:simpleVScomplex}
\end{figure*}

\begin{figure*}[h]
\centering
\includegraphics{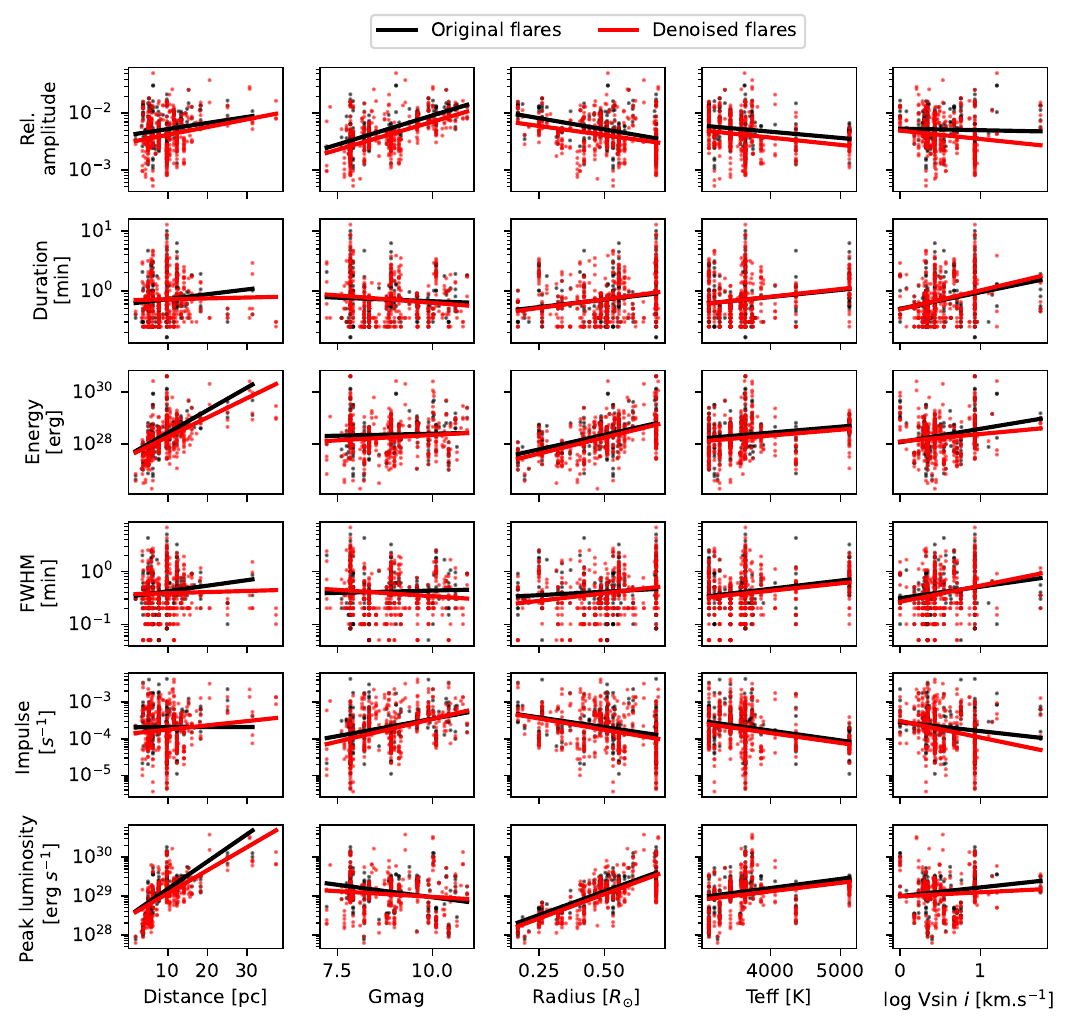}
\caption{Correlations between the flare parameters and the ones of its star, for flares recovered in original (black) and denoised (red) light curves. Linear fits are shown for indication.}
\label{annex:originalVSdenoised}
\end{figure*}

\end{appendix}

\end{document}